\documentclass[journal]{IEEEtran}

\usepackage{amsthm}
\usepackage{amsmath}
\usepackage[braket, qm]{qcircuit}
\usepackage{subfigure}
\usepackage{algorithm}
\usepackage{algpseudocode}
\usepackage{setspace}
\usepackage{threeparttable}
\usepackage{float}
\usepackage{graphicx}
\usepackage{multirow}
\usepackage{booktabs}
\usepackage{color}
\usepackage{diagbox}
\usepackage{lineno}
\usepackage{cite}
\usepackage{float}
\usepackage{stfloats}
\usepackage{algorithm,algpseudocode,float}
\usepackage{lipsum}
\usepackage{graphicx}
\usepackage{amsfonts}
\usepackage{bm}
\usepackage{mathrsfs}
\usepackage{graphicx}
\usepackage{float} 
\usepackage{subfigure}
\usepackage{diagbox}
\usepackage{mathrsfs}
\usepackage{tabularx}
\usepackage{arydshln}

\usepackage{amsthm}
\usepackage{amsmath}

\newtheorem{theorem}{Theorem}

\makeatletter

\renewcommand{\maketag@@@}[1]{\hbox{\m@th\normalsize\normalfont#1}}%

\makeatother

\begin{document}

	\title{Delegated variational quantum algorithms based on quantum homomorphic encryption}

\author{Qin Li, 
	Junyu Quan,
	Jinjing Shi, 
	{Shichao Zhang,
		and~Xuelong Li}
	
	\IEEEcompsocitemizethanks{\IEEEcompsocthanksitem  Qin Li and Junyu Quan are at the School of Computer Science, Xiangtan University, China, Jinjing Shi and Shichao Zhang are at the school of Computer Science and Engineering, Central South University, Xuelong Li is at the School of Artificial Intelligence, Northwestern Polytechnical University, China.}
	\thanks{Manuscript received January xxxx; revised xxxx.}}

\markboth{***,~Vol.~**, No.~**, January~2023}%
{Li \MakeLowercase{\textit{et al.}}: Delegated VQAs based on QHE}
	
	\maketitle
	
	\begin{abstract}
		Variational quantum algorithms (VQAs) are considered as one of the most promising candidates for achieving quantum advantages on quantum devices in the noisy intermediate-scale quantum (NISQ) era. They have been developed for numerous applications such as image processing and solving linear systems of equations. The application of VQAs can be greatly enlarged if users with limited quantum capabilities can run them on remote powerful quantum computers. But the private data of clients may be leaked to quantum servers in such a quantum cloud model. To solve the problem, a novel quantum homomorphic encryption (QHE) scheme which is client-friendly and suitable for VQAs is constructed for quantum servers to calculate encrypted data. Then delegated VQAs are proposed based on the given QHE scheme,  where the server can train the ansatz circuit using the client's data even without knowing the real input and the output of the client. Furthermore, a delegated variational quantum classifier to identify handwritten digit images is given as a specific example of delegated VQAs and simulated on the cloud platform of Original Quantum to show its feasibility.	
	\end{abstract}
	
	
	\section{Introduction}\label{sec:1}

\IEEEPARstart{Q}{uantum} computation can efficiently solve certain problems that are rather difficult with classical computation, such as factoring big integers \cite{shor1994algorithms}, simulating quantum systems \cite{lloyd1996universal}, and solving linear systems of equations \cite{harrow2009quantum}. Even though quantum computing has several benefits and applications \cite{LiEfficientqbcm,TruongRandom}, due to inherent limitations of quantum hardware, it is common to control a quantum system of over fifty but less than a few hundred qubits at present and thus lies in the NISQ era \cite{preskill2019quantum,zhang2017observation}. In the field of quantum machine learning \cite{ITOPXiao,MARTINGUERRERO2022457}, variational quantum algorithms (VQAs) are regarded as one important class of algorithms that can be realized in the NISQ era. They provide a general framework for solving practical problems such as quantum neural networks \cite{schuld2019quantum,beer2020training}, variational quantum classifier \cite{CabriClassifier} and variational Hamiltonian learning \cite{Vhashi} in the form of hybrid quantum-classical algorithms. They can be described as parametrized ansatz circuits which use classical optimizers to update the parameters for optimizing cost functions related to specific problems. 

In quantum networks, clients with limited quantum capabilities may upload their data to a remote quantum server to complete training tasks. In such a scenario, the server Bob needs to train a generic model with a delegated VQA by using private data from the user Alice who does not wish to expose her private data to other entities. In a secure delegated VQA where Alice inputs her private data and Bob provides the ansatz circuit, Bob should not obtain Alice's private data after implementing the protocol. In order to achieve this task, VQAs based on blind quantum computation (BQC) is proposed to complete variational secure cloud quantum computing\cite{PhysRevA.105.022603, quantumfederatedlearning}. However, BQC requires that the server should not know the input, output and algorithm of the user, so a malicious user can drive the server to perform the computation he wants instead of the training task. The server cannot detect the malicious behavior of the user during the computation and also cannot get the desired model. Besides, BQC usually needs large-scale entangled states and frequent interaction during the process of computation, which are very inefficient.

We observe that secure delegated VQAs can be realized better by using quantum homomorphic encryption (QHE) instead of BQC in two aspects. One is that QHE can enable quantum servers to perform calculations on encrypted data directly and make users get the expected results after decrypting the data returned by quantum servers. The other is that only one interaction between the server and the user is necessary.

In 2013, Liang gave definitions of QHE and quantum fully homomorphic encryption (QFHE) and constructed four symmetric QHE protocols and one symmetric QFHE protocol based on the quantum one-time pad \cite{liang2013symmetric}. In 2015, Liang proposed a QFHE protocol based on the universal set $\textsf{\{X,Y,Z,H,S,T,CNOT\}}$ \cite{liang2015quantum}. Broadbent and Jeffery gave two QHE schemes for the circuits with a limited number of non-Clifford gates such as \textsf{T}-gates \cite{broadbent2015quantum}. Later, Dulek et al. improved the protocol in Ref. \cite{broadbent2015quantum} and allowed it to implement polynomial-sized \textsf{T}-gates \cite{dulek2016quantum}. In 2018, Mahadev et al. proposed a QFHE scheme based on classical keys, in which a classical client is allowed to blindly delegate a quantum computation to a quantum server who cannot learn any information about the computation \cite{mahadev2020classical}. Several other QHE schemes have also been proposed based on different methods \cite{tan2016quantum,marshall2016continuous,alagic2017quantum,lai2017statistically,liu2022efficient}.

However, the existing QHE scheme can only implement a constant number of \textsf{T} gates, which is not enough to implement VQAs, such as Ref. \cite{broadbent2015quantum}, or the capabilities of the client is high, not only need to generate quantum states and implement \textsf{X,Z} gates, but also need to perform Bell measurements and $\textsf{P}^{\dagger}$ gate such as Ref. \cite{dulek2016quantum}. In addition, for Ref \cite{mahadev2020classical}, the capabilities of the client can be reduced to pure classical, but since its general gate set is \{Clifford + \textsf{Toffoli}\}, the implementation of \textsf{Toffoli} gate is much more difficult than \textsf{T} gate, which is also a heavy burden for the server. In this paper, we propose an efficient QHE scheme and then give a general framework for delegated VQAs based on the proposed QHE scheme. A specific example is also given and implemented on the cloud platform of Original Quantum. The main contributions of this paper can be summarized as follows.
\begin{itemize}
	\item A client-friendly QHE scheme suitable for constructing the general framework of VQAs is proposed, which can be served as the basis for distributed quantum privacy computing. In this QHE scheme, the client only needs to generate input qubits and implement \textsf{X} and \textsf{Z} gates, which are the minimum requirements when the input and output are quantum states. 
	\item A delegated variational quantum classifier used for identifying handwritten digit images is given as an example of delegated VQAs and simulated on the cloud platform of Original Quantum to demonstrate its feasibility. 
\end{itemize}

The rest part of the paper is organized as follows. Section \ref{sec:2} briefly introduces preliminaries related to QHE and VQAs. Section \ref{sec:3} reviews a typical QHE scheme, namely the  \textsf{TP} scheme in Ref. \cite{dulek2016quantum}. In section \ref{sec:4}, a novel QHE scheme is given. In section \ref{sec:5}, the delegated VQAs based on the given QHE scheme is proposed and an example of them is implemented on the cloud platform of Original Quantum in section \ref{sec:6}. The last section makes a conclusion.

\section{Preliminaries} \label{sec:2}
In this section, the definitions of classical homomorphic encryption (CHE) and QHE \cite{broadbent2015quantum,dulek2016quantum} are introduced. Besides, the basic knowledge of VQAs  \cite{cerezo2021variational,cojocaru2021possibility} is also given.

\subsection{Some definitions related to CHE and QHE}
A CHE scheme \textsf{HE} consists of four algorithms: key generation \textsf{HE.KeyGen}, encryption \textsf{HE.Enc}, evaluation \textsf{HE.Eval}, and decryption \textsf{HE.Dec}. With the application of \textsf{HE.KeyGen}, a public encryption key $pk$, an evaluation key $evk$, and a secret key $sk$ are generated, where the first two keys are public and the last one is only known to the client. The user Alice can encrypt the inputs $(x_1, \dots, x_l)$ with the public key $pk$ and send the ciphertext $(c_1, \dots, c_l)$ to the server Bob. Then, Bob evaluates the circuit $C$ with $evk$ on the ciphertext and returns the results back. Finally, Alice decrypts the results by the secret key $sk$ and obtains the output $C(x_1, \dots, x_l)$. The more formal definition of CHE is given in the following.
\newtheorem{definition}{Definition}
\begin{definition}
	A CHE scheme {\rm\textsf{HE}} consists of the following four algorithms: 
	
	\noindent\textbf{Key Generation.} {\rm\textsf{HE.KeyGen}}$(1^{\kappa}) \rightarrow (pk, sk, evk)$, where $\kappa \in \mathbb{N}$ is the security parameter, $1^{\kappa}$ is the input and three keys $pk$,  $sk$, and $evk$ are the output. \\
	\textbf{Encryption.} {\rm\textsf{HE.Enc$_{pk}$}}$(x)$ $\rightarrow$ $c$, which maps one-bit message $x\in \{0,1\}$ to a ciphertext $c$ with $pk$. \\
	\textbf{Homomorphic Evaluation.} {\rm\textsf{HE.Eval$^{C}_{evk}$}}$(c_{1},...,c_{l})$ $\rightarrow$ $c^{'}$, which implements the evaluation circuit $C$ on the ciphertext $(c_1, \dots, c_l)$ with $evk$ to get $c^{'}$ .     \\
	\textbf{Decryption.} {\rm\textsf{HE.Dec$_{sk}$}}$(c^{'})$ $\rightarrow$ $x^{'}$, which maps the result $c^{'}$ for the ciphertext $(c_1, \dots, c_l)$ to $x^{'}$ for the plaintext $(x_1, \dots, x_l)$ with $sk$.
\end{definition}

Similarly, in a QHE scheme \textsf{QHE}, Alice first employs \textsf{QHE.KeyGen} to obtain a classical public key $pk$, a classical secret key $sk$, and a quantum evaluation key $\rho_{evk}$. She implements the encryption operation \textsf{QHE.Enc} on the inputs with $pk$ and then sends the ciphertext to Bob. After Bob applies \textsf{QHE.Eval} on the ciphertext with $\rho_{evk}$, he sends the result back to Alice. Finally, Alice carries out the decryption operation \textsf{QHE.Dec} on the calculation result that Bob offered with $sk$ to obtain the real output. The more specific definition of \textsf{QHE} is described as follows.
\begin{definition}
	A QHE scheme {\rm \textsf{QHE}} is made up of the following four algorithms: 
	
	\noindent\textbf{Key Generation.} {\rm\textsf{QHE.KeyGen}} $(1^{\kappa})$ $\rightarrow$ $(pk$, $sk$, $\rho_{evk})$, 
	where $\kappa \in \mathbb{N}$ is the security parameter, $1^{\kappa}$ is the input and three keys $pk$,  $sk$, and $\rho_{evk}$ are the output.
	\\
	\textbf{Encryption.} {\rm\textsf{QHE.Enc$_{pk}$}}$(\rho)$ $\rightarrow$ $\sigma$, which maps an input state $\rho$ to a cipherstate $\sigma$ with $pk$.  \\
	\textbf{Homomorphic Evaluation.} {\rm\textsf{QHE.Eval$^\textsf{C}_{\rho_{evk}}$}}$(\sigma)$ $\rightarrow$ $\sigma^{'}$, which changes the cipherstate $\sigma$ to $\sigma'$ according to $\rho_{evk}$.    \\
	\textbf{Decryption.} {\rm\textsf{QHE.Dec$_{sk}$}} $(\sigma^{'}) $$\rightarrow$ $\rho^{'}$, which maps a single state $\sigma^{'}$ to $\rho^{'}$, which is the calculation result of the real input $\rho$.
\end{definition}

\begin{figure}
	\centering
	\includegraphics[width=0.5\textwidth]{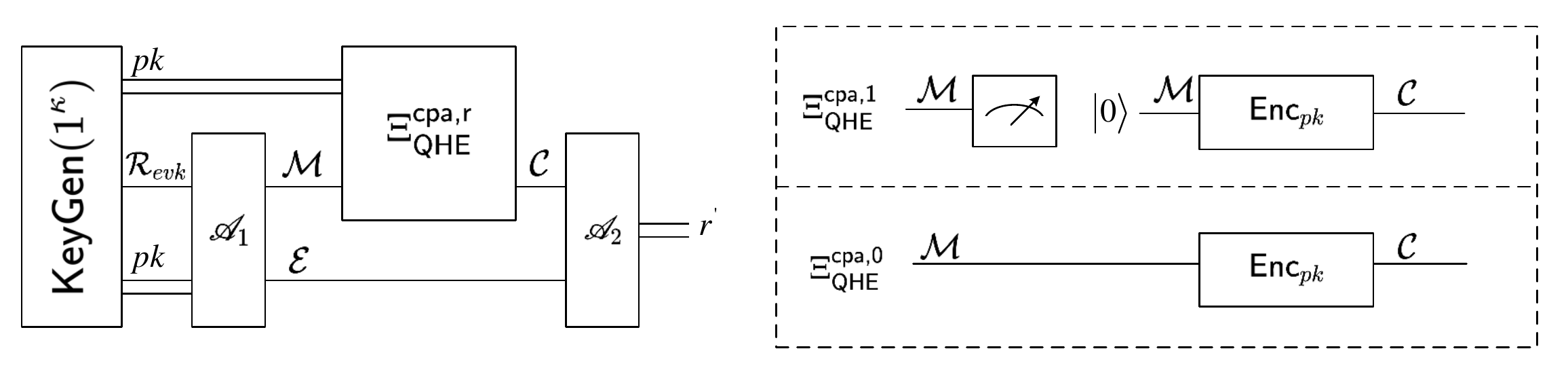}
	\caption{The game model of quantum indistinguishability under CPA} \label{zm1}
\end{figure}

As for the security of a QHE scheme, it should satisfy indistinguishability under chosen-plaintext attacks (q-IND-CPA) in quantum polynomial (QPT) time \cite{dulek2016quantum}. Hence, a QHE scheme is said to be q-IND-CPA secure if for any QPT adversary $\mathscr{A} = (\mathscr{A}_1,\mathscr{A}_2)$ there exists a negligible function satisfying
\begin{equation}
	Pr[\textsf{PubK}^{\textsf{cpa}}_{\mathscr{A},\textsf{QHE}}(\kappa)=1]\leq \frac{1}{2}+\mathsf{negl}(\kappa).
\end{equation}
where $\textsf{PubK}^{\textsf{cpa}}_{\mathscr{A},\textsf{QHE}}$ is a model of quantum indistinguishability under CPA as shown in Fig. \ref{zm1}.

\begin{definition}[Quantum indistinguishability under CPA]
	The game model of quantum indistinguishability under chosen-plaintext attack (IND-CPA) {\rm $\textsf{PubK}^{cpa}_{\mathscr{A,\textsf{QHE}}}(\kappa)$} for a {\rm\textsf{QHE}} scheme and a {\rm QPT} adversary $\mathscr{A} = (\mathscr{A}_1,\mathscr{A}_2)$ is defined as
	
	1. The challenger runs {\rm \textsf{QHE.KeyGen($1^{\kappa}$)}} $\rightarrow$ $(pk, sk, \rho_{evk}$). 
	
	2. The challenger sends $(pk, \rho_{evk})$ to $\mathscr{A}_1$. Then $\mathscr{A}_1$ outputs a quantum state in $\mathcal{M} \otimes \mathcal{E}$, where $\mathcal{M}$ is the message space and $\mathcal{E}$ is an arbitrary state related to the environment.
	
	3. For $r\in\{0,1\}$, let {\rm $\Xi_{\textsf{QHE}}^{\textsf{cpa},r}$: $D(\mathcal{M}) \rightarrow D(\mathcal{C})$} be {\rm $\Xi_{\textsf{QHE}}^{\textsf{cpa},0}(\rho)=\textsf{QHE.Enc}_{pk}(|0\rangle \langle0|)$} and {\rm $\Xi_{\textsf{QHE}}^{\textsf{cpa},1}(\rho)=\textsf{QHE.Enc}_{pk}(\rho)$}. A random bit $r\in\{0,1\}$ is chosen and {\rm $\Xi_{\textsf{QHE}}^{\textsf{cpa},r}$} is applied to the state in $\mathcal{M}$.
	
	4. $\mathscr{A}_2$ obtains the state in $\mathcal{C}\otimes\mathcal{E}$ and outputs a bit $r^{\prime}$. 
	
	5. The output of the game is defined to be 1 if $r^{\prime} = r$ and 0 otherwise. If $r = r^{\prime}$, $\mathscr{A}_2$ wins the game.
\end{definition}
\subsection{VQAs}

\begin{figure}
	\centering
	\includegraphics[width=0.5\textwidth]{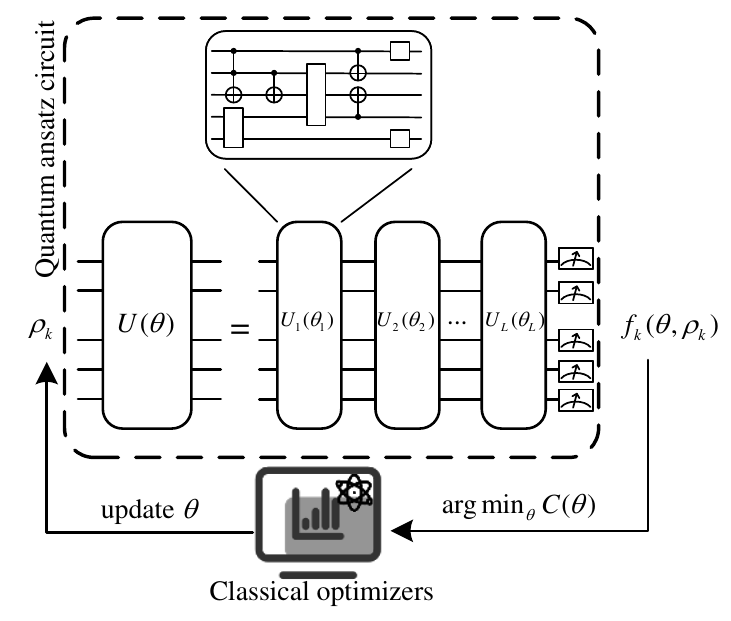}
	\caption{The schematic diagram of VQAs and the construction of a quantum ansatz circuit} \label{VQA}
\end{figure}

VQAs are hybrid quantum-classical algorithms which can be used to solve a variety of problems. As shown in Fig. \ref{VQA}, VQAs use a quantum computer to estimate the cost function $C(\theta)$ as a solution to a required task and it can be defined as  
\begin{equation}
	\label{eq:1}
	\begin{split}
		C(\theta) = \sum_k f_k({\rm Tr}[O_kU(\theta)\rho_kU^\dagger(\theta)]),
	\end{split}
\end{equation}
where $\{f_k\}$ is a set of some functions, ${\rho_k}$ are input states and $O_k$ are observables such as Pauli operators $P_{i} \in \{\textsf{I,X,Y,Z}\}^{\otimes n}$, $U(\theta)$ is the parametrized ansatz quantum circuit, and $\theta$ is a variational parameter which can be trained by the classical optimizer to solve the optimization task
\begin{equation}
	\theta^* = \mathop{\arg\min}\limits_{\theta}C(\theta).
\end{equation}
And a multi-layer layout ansatz $U(\theta)$ can be expressed as 
\begin{equation}
	U(\theta)=\prod_{r=1}^{R} U_{r}(\theta_r),
\end{equation}
where $U_r(\theta_r)=\prod_m e^{-i\theta_{m}H_m}W_m$, $W_m$ is an unparametrized unitary operator and $H_m$ is a Hermitian operator. Then VQAs are used to train the parameters $\theta$ iteratively to minimize the cost function $C(\theta)$ according to the classical optimizer. At the $t$-th iteration, the updating rule is $\theta^{(t+1)} = \theta^{(t)}-\chi\frac{\partial C(\theta)}{\partial \theta}$, where $\chi$ is the learning rate and the partial derivative of $C(\theta)$ with respect to $\theta$ is defined as
\begin{equation}
	\begin{split}
		\frac{ \partial C }{ \partial \theta } = & \sum_k \frac{1}{2{\rm sin} \alpha} ({\rm Tr}[O_k U^{\dagger}(\theta_+)\rho_k]U(\theta_+)] \\&-{\rm Tr}[O_k U^{\dagger}(\theta_-)\rho_k]U(\theta_-)]),	
	\end{split}
\end{equation}
with $\theta_\pm = \theta \pm \alpha e_l$ for any real number $\alpha$ and $e_l \in \{0,1\}$ is a vector. 

\section{Review of the \textsf{TP} scheme \cite{dulek2016quantum}} \label{sec:3}

In this part, a typical QHE scheme called \textsf{TP} scheme is briefly reviewed \cite{dulek2016quantum}. As well known, Clifford gates \textsf{\{X, Z, P, CNOT, H\}} and any one kind of non-Clifford gates such as \textsf{T} gate can be used to construct a universal gate set for quantum computation. In \textsf{TP} scheme, such gates \textsf{\{X, Z, P, CNOT, H, T\}} can be applied to encrypted states and the output states also can be decrypted to obtain the results about the original states. The main steps are given in the following.

Firstly, the client employs the quantum one-time pad to encrypt each single-qubit state $|\psi\rangle $ to obtain
\begin{equation}
	|\psi\rangle_{encrypted}=\textsf{X}^{a}\textsf{Z}^{b}|\psi\rangle,
\end{equation}
where $a,b \in \{0,1\}$ are secret key bits randomly generated by the client. Then she sends $|\psi\rangle_{encrypted}$ to the server and the server performs quantum gates in the set \textsf{\{X, Z, P, CNOT, H, T\}} on them to achieve the specific computational task. Since the non-Clifford \textsf{T} gate does not commute with the Pauli \textsf{X} gate and $\textsf{TX}^{a}\textsf{Z}^{b} = \textsf{P}^{a}\textsf{X}^{a}\textsf{Z}^{b}\textsf{T}$, the client has to correct the by-product $\textsf{P}$ by telling the server the value of $a$, which results in the secret key bit being revealed. The \textsf{TP} scheme \cite{dulek2016quantum} can solve this problem by using a \textsf{T} gate gadget. The key idea is that an inverse phase gate can be applied on the qubit $\textsf{X}^{a^\prime} \textsf{Z}^{b^\prime} \textsf{P}|\psi\rangle$ by using $(\textsf{P}^\dagger \otimes \textsf{I})|\Phi^{+}\rangle$ to teleport a qubit $\textsf{X}^a \textsf{Z}^b |\psi\rangle$, where $|\Phi^{+}\rangle = \frac{1}{\sqrt{2}}(|00\rangle + |11\rangle)$ and the new Pauli corrections $a^\prime, b^\prime$ depend on $a,b$ and the outcome of the Bell measurement. 

The gadget consists of a classical part and a quantum part. Based on a secret key $sk$ for a classical \textsf{HE}, the classical part $g(sk)$ is defined as
\begin{equation}
	g(sk) = (\{(s_1, t_1),(s_2, t_2), \dots ,(s_m, t_m)\}, p, sk),
\end{equation}
where $m$ relies on a security parameter $\kappa$, $p \in \{0, 1\}^{m}$ is a string of $m$ bits, and ${(s_1, t_1),(s_2, t_2), \dots ,(s_m, t_m)}$ are disjoint pairs in $\{1, 2, \dots , 2m\}$. The corresponding quantum part consists of $2m$ qubits and is defined as
\begin{equation}
	\label{eq:11}
	\gamma_{x, z}(g(s k))=\prod_{i=1}^{m} \textsf{X}^{x[i]} \textsf{Z}^{z[i]}(\textsf{P}^{\dagger})^{p[i]}|\Phi^{+}\rangle \langle\Phi^{+}|_{s_{i} t_{i}} \textsf{P}^{p[i]} \textsf{Z}^{z[i]} \textsf{X}^{x[i]},
\end{equation}
where $x, z \in \{0, 1\}^m$ are the Pauli key strings and $x[i], z[i]$, and $p[i]$ are the $i$-th bits of the strings $x$, $z$, and $p$, respectively. Therefore, the entire gadget is given by
\begin{equation}
	\begin{split}
		\Gamma_{p k^{\prime}}(s k)=&[\textsf{HE.Enc}_{p k^{\prime}}(g(s k)),\\& \frac{1}{2^{2 m}} \sum_{x,z\in\{0,1\}} \textsf{HE.Enc}_{p k^{\prime}}(x, z)] || \gamma_{x, z}(g(s k)) .
	\end{split}
\end{equation}

To utilize the gadget, the server needs to perform a Bell measurement between the gadget qubit and an input qubit and make Pauli operations on the output qubit based on the measurement result. The order of measurements is decided by a classical algorithm \textsf{GenMeasurement($\widetilde{a}$)} which produces a list $M$ which contains $m$ disjoint pairs of elements in $\{0, 1, 2,..., 2m\}$, where the label $1$ to $2m$ refer to the gadget qubits and 0 is the input qubit. After all the Bell measurements have been performed with the order of measurement in $M$, the remaining single qubit is the output qubit. 

\section{The proposed QHE scheme} \label{sec:4}
In this part, a client-friendly QHE scheme namely $\textsf{QHE}_\textsf{CC}$ is proposed. Its security is analyzed and comparisons with other similar QHE protocols are also made.

\subsection{The proposed {\rm$\textsf{QHE}_\textsf{CC}$} scheme}

The proposed $\textsf{QHE}_\textsf{CC}$ scheme is an extension of the \textsf{TP} scheme and the difference mainly lies in the ways of generating gadgets in the key generation algorithm. In the \textsf{TP} scheme \cite{dulek2016quantum}, the client needs some quantum ability such as generating EPR-pairs, performing $\textsf{P}^{\dagger}$ gates and Bell measurements to construct gadgets for removing byproducts reduced by \textsf{T} gates. However, in the proposed $\textsf{QHE}_\textsf{CC}$ scheme, a novel algorithm \textsf{GenGadget} is proposed which can allow a purely classical client to generate the gadget securely with a quantum server. If the server honestly follows the algorithm, the client could generate the correct gadget. Otherwise, if the server is malicious, he cannot obtain any useful information except the number of \textsf{T} gates. We use the $\textsf{CC-RSP}_{\theta}$ \cite{cojocaru2021possibility} to construct the algorithm \textsf{GenGadget}, which provides a way to generate random remotely single qubits $|+_{\theta}\rangle$ defined as  
\begin{equation}
	|+_{\theta}\rangle = \frac{1}{\sqrt{2}}(|0\rangle + e^{i \theta}|1\rangle), \theta \in \{0,\frac{\pi}{4},...,\frac{7\pi}{4}\}.
\end{equation}
But in the $\textsf{QHE}_\textsf{CC}$ scheme, the client only needs to generate $|+_{\theta}\rangle$ where $\theta\in\{0, \frac{\pi}{2},\pi,\frac{3\pi}{2}\}$, and then the server perform a fixed coupling operation $(\textsf{I} \otimes \textsf{H}) \textsf{CZ}$ on these qubits to generate different entangled states, hence $\textsf{CC-RSP}_{\theta}$ needs some modifications. For example, suppose that $\textsf{CC-RSP}_{\theta}$ enables the client to generate a single-qubit $\{|0\rangle + e^{i \theta} |1\rangle \}$ remotely where $\theta\in\{0, \frac{\pi}{2},\pi,\frac{3\pi}{2}\}$, corresponding to the state

\begin{small}
	\begin{equation}
		\begin{split}
			|+_{0}\rangle &= |+\rangle = \frac{1}{\sqrt{2}}(|0\rangle+|1\rangle),  |+_{\frac{\pi}{2}}\rangle = \textsf{P}|+\rangle = \frac{1}{\sqrt{2}}(|0\rangle+i|1\rangle), \\
			|+_{\frac{3\pi}{2}}\rangle &= \textsf{P}^{\dagger}|+\rangle = \frac{1}{\sqrt{2}}(|0\rangle-i|1\rangle),  |+_{\pi}\rangle = |-\rangle = \frac{1}{\sqrt{2}}(|0\rangle-|1\rangle).
		\end{split}
	\end{equation}
\end{small}

Any two of these qubits are selected and a fixed coupling operation $(\textsf{I} \otimes \textsf{H}) \textsf{CZ}$ is performed on them to obtain
\begin{equation}
	\label{eq:15}
	\begin{split}
		\textsf{CZ}(|+\rangle \otimes |+\rangle)&\stackrel{\textsf{I} \otimes \textsf{H}} {\longrightarrow}  \textsf{X}^{0}\textsf{Z}^{0}|\Phi^{+}\rangle, 
		\textsf{CZ}(|+\rangle \otimes |-\rangle)\stackrel{\textsf{I} \otimes \textsf{H}}{\longrightarrow}  \textsf{X}^{0}\textsf{Z}^{1}|\Phi^{+}\rangle, \\
		\textsf{CZ}(|-\rangle \otimes |+\rangle)& \stackrel{\textsf{I} \otimes \textsf{H}} {\longrightarrow}   \textsf{X}^{1}\textsf{Z}^{0}|\Phi^{+}\rangle,  
		\textsf{CZ}(|-\rangle \otimes |-\rangle)\stackrel{\textsf{I} \otimes \textsf{H}}{\longrightarrow} \textsf{X}^{1}\textsf{Z}^{1}|\Phi^{+}\rangle, \\
		&\textsf{CZ}(|+\rangle \otimes |+_{\frac{3\pi}{2}}\rangle)\stackrel{\textsf{I} \otimes \textsf{H}}{\longrightarrow}   \textsf{X}^{0}\textsf{Z}^{0}P^{\dagger}|\Phi^{+}\rangle,\\
		&\textsf{CZ}(|+\rangle \otimes |+_{\frac{\pi}{2}}\rangle)\stackrel{\textsf{I} \otimes \textsf{H}}{\longrightarrow}  \textsf{X}^{0}\textsf{Z}^{1}P^{\dagger}|\Phi^{+}\rangle,\\  
		&\textsf{CZ}(|-\rangle \otimes |+_{\frac{3\pi}{2}}\rangle)\stackrel{\textsf{I} \otimes \textsf{H}}{\longrightarrow}   \textsf{X}^{1}\textsf{Z}^{0}P^{\dagger}|\Phi^{+}\rangle,\\
		&\textsf{CZ}(|-\rangle \otimes |+_{\frac{\pi}{2}}\rangle)\stackrel{\textsf{I} \otimes \textsf{H}}{\longrightarrow}  \textsf{X}^{1}\textsf{Z}^{1}P^{\dagger}|\Phi^{+}\rangle.  \\ 
	\end{split}
\end{equation}

According to Eq. (\ref{eq:15}), Bob can obtain the quantum state as shown in Eq. (\ref{eq:11}) by following Alice's instructions, which is the quantum part of the gadget.

The proposed $\textsf{QHE}_\textsf{CC}$ scheme also contains four algorithms: key generation, encryption, homomorphic evaluation and decryption. Next, the steps of the $\textsf{QHE}_\textsf{CC}$ scheme are given as follows.

\subsubsection{Key Generation}

Assume the client Alice wants to execute the quantum computation containing $L$ \textsf{T} gates with a security parameter $\kappa$. Then the key generation algorithm \textsf{$\textsf{QHE}_\textsf{CC}$.\textsf{KeyGen}($1^{\kappa},1^{L}$)} is defined as:

1. For $i\in[0,L] $, Alice needs to perform $\textsf{HE.KeyGen}(1^{\kappa})$ to generate a series of classical secret keys $(sk_{i})^{L}_{i=0}$ and public keys $(pk_{i})^{L}_{i=0}$, and classical keys $(evk_{i})^{L}_{i=0}$ for quantum evaluation. 

2. For $i\in[0,L-1]$, Alice repeats Algorithm 1 \textsf{GenGadget} to create a \textsf{T} gate gadget $\Gamma_{pk_{i+1}}(sk_i)$ in Bob's hand. Then, according to $(evk_{i})^{L}_{i=0}$, Bob could get the quantum evaluation key as
\begin{equation}
	\rho_{evk_i}=\bigotimes^{L-1}_{i=0}(\Gamma_{pk_{i+1}}(s k_{i})\otimes|evk_{i}\rangle\langle evk_{i}|).
\end{equation}
Note that, $pk_0$ is used to encrypt the Pauli keys $a, b$ in the encryption algorithm and the remaining $L$ $pk_{i+1}$ is used to encrypt the gadget $\Gamma_{pk_{i+1}}$, where $i\in[0,L-1]$.
\begin{algorithm*}[htb]
	\caption{\textsf{GenGadget}}
	\label{alg:A}
	\noindent{\bf Requirements:}
	The client Alice chooses a family of one-way trapdoor functions namely $ \mathcal{F} = \{ f_k: \{0,1\}^n  \rightarrow \{0,1\}^\mu \} $, which should be quantum-secure, two-regular and collision resistant \cite{cojocaru2021possibility}. In addition, according to $k$ which is public, Alice also chooses her own private trapdoor information $t_k$.\\
	\noindent{\bf Input:}
	Alice chooses a string $\alpha = (\alpha_1,...,\alpha_{n-1})$ randomly, where $\alpha_i \in \{0,1\}$.	\\
	\noindent For $1 \leq i \leq 2m$:
	\begin{algorithmic}
		\State Step 1. {Alice asks Bob to prepare two registers in states \ $\otimes^n \textsf{H} |0\rangle$ and $|0\rangle^{\otimes \mu}$, respectively. Then Bob applies the controlled-unitary operation $U_{f_k}$ on the two registers, where the first register stores the control qubits, the second register stores the target qubits, and the funtion $f_k$ is defined as
			\begin{equation}
				\forall f_k:A\rightarrow B, \exists x\in A, y\in B, U_{f_k}|x\rangle|y\rangle=|x\rangle|y\oplus f_{k}(x)\rangle.
			\end{equation}
			Therefore, the state after applying the $U_{f_k}$ gate is
			\begin{equation}
				\begin{split}
					U_{f_k}(|+\rangle^{n} \otimes |0\rangle^{\mu}) &= |+\rangle^{n} \otimes |0\oplus f_{k}(|+\rangle) \rangle ^{\mu}\\
					&=\sum\nolimits_{x}|x\rangle^{n}  \otimes |f_{k}(x)\rangle^{\mu}, x\in\{0,1\}.
				\end{split}
		\end{equation}}
		\State Step 2. {Bob measures all qubits in the second register in the $\textsf{Z}$ basis and returns the measurement result $y$ to Alice. Because $f_k$ is a two regular function and $y=f_k(x) = f_k(x^{'}) $, it is not difficult to deduce that the state in the first register collapses to $(|x\rangle+|x^{'} \rangle)$ after the  measurement. Note that $f_k$ is also a collision resistant function, therefore $x\neq x^{'}$.}
		\State Step 3. {Bob measures all but the last qubits in the first register in the basis $\{|0\rangle \pm e^{\alpha_i \pi/2} |1\rangle  \}$ and returns the measurement results $b = (b_1, \dots, b_{n-1})$ to Alice. }
		\State Step 4. {Alice can easily computes $(x, x^{'})=\textsf{Inv}_{f_k}(t_k,y)$ according to the inversion algorithm \textsf{Inv} of $f_k$, because she has the trapdoor information $t_{k}$. Then she checks whether the value of $n$-th bit $x$ and $x^{'}$ are equal. If they are not equal, Alice could recover the classical description of the server’s state as
			\begin{equation}
				\label{eq:19}
				\theta_i= \frac{\pi}{2}(-1)^{x_n}(\sum_{i=1}^{n-1}(x_i - x_i^{'})(2b_i+\alpha_i)) \ mod\ 4.
			\end{equation}
			Otherwise, Alice terminates the algorithm and returns to Step 1.}
		\State Step 5. {Alice and Bob repeat steps 1 to 4 until $2m$ qubits  are generated in Bob's hand and the state of each qubit is $|+_{\theta\rangle_i}$.  }
	\end{algorithmic}
	\noindent{\bf Output:}  For $2m$ qubits, Alice divides them into two sets $\{(s_1, t_1), (s_2, t_2), \dots, (s_m, t_m)\}$ and sends related classical information $g(sk_i)$ to Bob. Bob performs $(\textsf{I} \otimes \textsf{H}) \textsf{CZ}$ as Alice required on these qubits and according to Eq (\ref{eq:15}), the quantum state becomes
	\begin{equation}
		\gamma_{x, z}=\prod_{i=1}^{m} \textsf{X}^{x[i]} \textsf{Z}^{z[i]}(\textsf{P}^{\dagger})^{p[i]}|\Phi^{+}\rangle \langle\Phi^{+}|_{s_{i} t_{i}} \textsf{P}^{p[i]} \textsf{Z}^{z[i]} \textsf{Z}^{x[i]},
	\end{equation}
	where $x[i],z[i] \in \{0,1\}$, and $p[i]$ can be deduced by Alice based on Eq. (\ref{eq:15}). Then Alice encrypts the information $g(sk_i) = (\{(s_1, t_1),(s_2, t_2), \dots ,(s_m, t_m)\}, p, sk_i)$ with public key $pk_{i+1}$. Note that, the length of $g(sk_i)$ determined by the choice of \textsf{HE} and the security parameter $\kappa$, Bob cannot deduce the value of $p$ and $sk_i$ through $g(sk_i)$ and $\{(s_1, t_1), (s_2, t_2), \dots, (s_m, t_m)\}$. Then the output of the entire gadget is described as
	\begin{equation}
		\Gamma_{pk_{i+1}}(sk_i)=[\textsf{HE.Enc}_{pk_{i+1}}(g(sk_i)),\frac{1}{2^{2 m}} \sum_{x,z\in\{0,1\}} \textsf{HE.Enc}_{pk_{i+1}}(x, z)] || \gamma_{x, z} .
	\end{equation}
\end{algorithm*}

\subsubsection{Encryption}

Alice encrypts each single input qubit $|\psi_i\rangle$ with $ \textsf{X}^{a_i} \textsf{Z}^{b_i}$, where $(a_{i},b_{i}) \in \{0,1\}$ are quantum one-time-pad keys and they should be encrypted by the first public key $pk_0$ and sent to Bob. Therefore, the encrypted classical-quantum state can be described as 
\begin{equation}
	\begin{split}
		\textsf{QHE}_{\textsf{CC}}.\textsf{Enc}_{pk0}(|\psi_i\rangle\langle\psi_i|) =&\sum_{i=1}^{n} (\textsf{HE.Enc}_{p k_{0}}(a_i,b_i) \\& \otimes  \frac{1}{4} \textsf{X}^{a_i} \textsf{Z}^{b_i} |\psi_i \rangle \langle \psi_i| \textsf{Z}^{b_i} \textsf{X}^{a_i}) .
	\end{split}
\end{equation}

\subsubsection{Homomorphic evaluation}

Bob applies unitary operations $U_r\in\{U_1, U_2, \dots, U_R\} $ on encrypted input states in the form of 
\begin{equation}
	(\textsf{X}^{a_{1}}  \textsf{Z}^{b_{1}} \otimes \cdots \otimes \textsf{X}^{a_{n}}  \textsf{Z}^{b_{n}}) \rho\left(\textsf{X}^{a_{1}}  \textsf{Z}^{b_{1}} \otimes \cdots \otimes \textsf{X}^{a_{n}}  \textsf{Z}^{b_{n}}\right)
\end{equation}
received from Alice, where $\rho=|\psi_i \rangle \langle \psi_i|^{\otimes n}$ and $(U_i)_{i=1}^{R} \in G= \{\textsf{X, Z, H, P, CNOT, T}\}$. There are two cases for the evaluation.

\begin{figure}[H]
	\centering
	\includegraphics[width=0.45\textwidth]{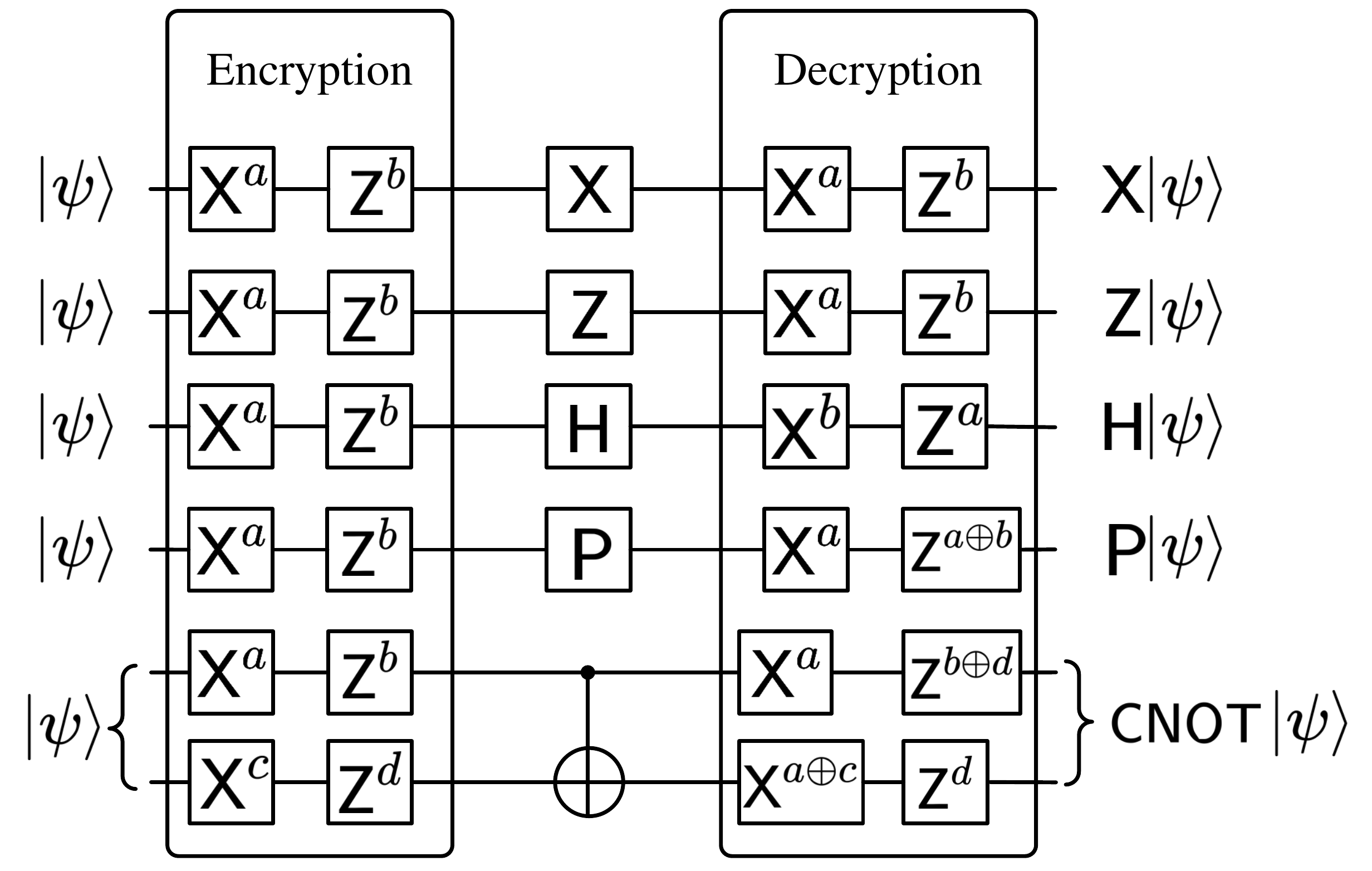}
	\caption{The encryption, homomorphic evaluation and decryption for quantum gates in the universal set $ G= \{\textsf{X, Z, H, P, CNOT, T}\}$.} \label{fig3}
\end{figure}

(i) If $U_r = \{\textsf{X, Z, H, P, CNOT}\}$, the gate $U_r$ is simply applied to the encrypted qubit as shown in Fig. \ref{fig3}, as $U_r$ is a Clifford gate and commutes with the Pauli group.

\begin{figure}
	\centering
	\includegraphics[width=0.47\textwidth]{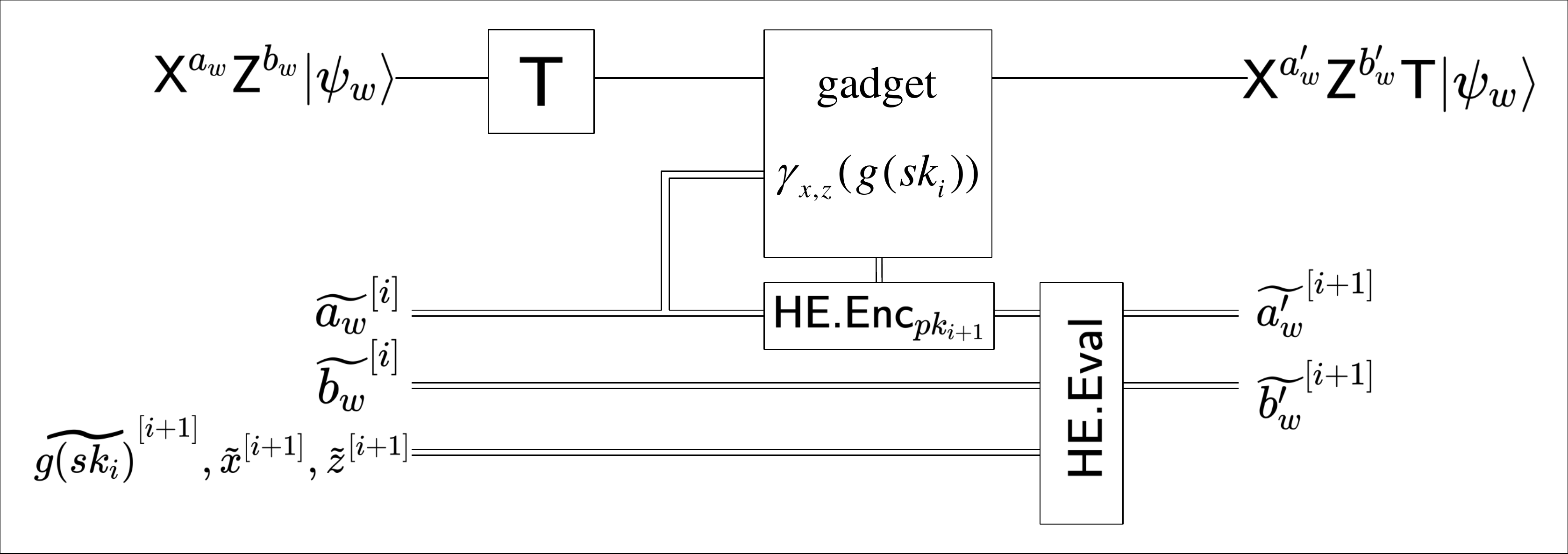}
	\caption{The homomorphic evaluation of \textsf{T} gate. The gadget is executed after the $(i+1)$th \textsf{T} gate. Then, Bob uses the classical algorithm \textsf{HE.Eval} to evaluates the new keys $\widetilde{a^{\prime}}^{[i+1]}_{w}$ and $\widetilde{b^{\prime}}^{[i+1]}_{w}$.  } \label{fig4}
\end{figure}

(ii) If $U_r = \textsf{T}$ and suppose it is on the $w$-th wire and the $i$-th \textsf{T} gate, Bob performs a \textsf{T} gate and the state becomes
\begin{equation}
	(\textsf{P}^{a_w}\textsf{X}^{a_w}\textsf{Z}^{b_w}\textsf{T})|\psi_w \rangle \langle \psi_w|(\textsf{T}^{\dagger}\textsf{X}^{a_w}\textsf{Z}^{b_w}(\textsf{P}^\dagger)^{a_w}).
\end{equation}
In order to remove the possible byproduct $\textsf{P}$, one gadget $\Gamma_{pk_{i+1}}(sk_i)$ according to the evaluation key is used. Based on the measurement sequence $M$ generated by $\textsf{GenMeasurement}(\widetilde{a}^{[i]}_{w})$ where $\widetilde{a}^{[i]}_{w}$ is classical information encrypted by $pk_{i}$, Bob makes the Bell measurement on $\textsf{P}^{a_w}\textsf{X}^{a_w}\textsf{Z}^{b_w}\textsf{T}|\psi_w \rangle$ and the pairs of gadgets. As shown in Fig. \ref{fig4}, in terms of the measurement results and the gadget's classical information $\widetilde{g(sk_i)}^{[i+1]}$ encrypted using $pk_{i+1}$, Bob homomorphically computes the new keys $\widetilde{a^{\prime}}^{[i+1]}_{w}$ and $\widetilde{b^{\prime}}^{[i+1]}_{w}$.

\subsubsection{Decryption}

Suppose that the state returned to Alice after the calculation is 

\begin{small}
	\begin{equation}
		(\textsf{X}^{a^{\prime}_{1}}  \textsf{Z}^{b^{\prime}_{1}} \otimes \cdots \otimes \textsf{X}^{a^{\prime}_{n}}  \textsf{Z}^{b^{\prime}_{n}}) U\rho U^{\dagger}(\textsf{X}^{a^{\prime}_{1}}  \textsf{Z}^{b^{\prime}_{1}} \otimes \cdots \otimes \textsf{X}^{a^{\prime}_{n}}  \textsf{Z}^{b^{\prime}_{n}}).
	\end{equation}
\end{small}

Bob also sends back to Alice the last updated classical key $(\widetilde{a^{\prime}_i},\widetilde{b^{\prime}_i})$. Alice can obtain $a^{\prime}_{i}$ via $\textsf{HE.Dec}_{sk}(\widetilde{a^{\prime}_i})$ and $b^{\prime}_{i}$ via $\textsf{HE.Dec}_{sk}(\widetilde{b^{\prime}_{i}})$. Then, she performs the gate $\textsf{X}^{a^{\prime}_{i}}$ $\textsf{Z}^{b^{\prime}_{i}}$ on each qubit to get the desired state.

\subsection{Security analysis and comparisons}

\begin{table*}[t]
	\tabcolsep 2pt
	\footnotesize
	\caption{Comparisons between the proposed protocol $\textsf{QHE}_{\textsf{cc}}$ and other QHE protocols}
	\label{tab4}
	\begin{tabular}{p{3.5cm}p{7.5cm}p{3cm}p{2cm}p{1.5cm}}
		\hline
		& Capabilities of clients & The number of T gates & Gate set & Security  \\ \hline
		\textsf{CL} scheme \cite{broadbent2015quantum}  & Performing \textsf{X,Z} gates, generating quantum input states  & Constant  & Clifford  & q-IND-CPA  \\
		
		\textsf{AUX} scheme \cite{broadbent2015quantum}  & Performing \textsf{X,Z} gates, generating ancillary states and quantum input states  & Constant  & Clifford +\textsf{T}  & q-IND-CPA  \\
		
		\textsf{TP} scheme \cite{dulek2016quantum}&  Making Bell measurements, performing $\textsf{X,Z,P}^{\dagger}$ gates and generating quantum input states & Polynomial & Clifford +\textsf{T}  & q-IND-CPA   \\
		
		Encrypted \textsf{CNOT} scheme \cite{mahadev2020classical} & Pure classical capabilities & N/A & Clifford +\textsf{Toffoli}  &  q-IND-CPA  \\
		
		The proposed $\textsf{QHE}_\textsf{cc} $ scheme &  Performing \textsf{X,Z} gates, generating quantum input states & Polynomial  &  Clifford +\textsf{T}  & q-IND-CPA \\
		
		\hline
	\end{tabular}
\end{table*}
In this part, the proposed $\textsf{QHE}_{\textsf{cc}}$ scheme is shown to satisfy q-IND-CPA security and made comparisons with similar QHE schemes. The $\textsf{QHE}_{\textsf{cc}}$ scheme can be regarded as an extension of $\textsf{TP}$ scheme to some extent, and the main difference is that the client runs \textsf{GenGadget} to generate the gadget, where the capabilities of Alice could be reduced to be classical. Therefore, we first show the security of the algorithm \textsf{GenGadget} and then show the proposed scheme satisfies q-IND-CPA security .

\begin{theorem} \label{theorem:1}
	In the algorithm {\rm\textsf{GenGadget}}, for any QPT adversaries $\mathscr{A}$, he cannot get any useful information about the quantum part of the gadget.
\end{theorem}

\proof In the algorithm {\rm\textsf{GenGadget}}, the client generates $|+_\theta\rangle$ where $\theta\in\{0,\frac{\pi}{2},\pi,\frac{3\pi}{2}\}$ by using $\textsf{CC-RSP}_{\theta}$, which has been proven that for any QPT adversaries, he cannot get the correct $\theta$ in the client's hand with the probability less than $\frac{1}{2}+\textsf{negl(n)}$ \cite{cojocaru2021possibility}. In the following, clients are considered as challengers, while servers are considered as adversaries. Therefore, the algorithm {\rm\textsf{GenGadget}} can be simplified as that a challenger chooses a classical bit $c\in\{0,1\}$ randomly, then she follows the steps of algorithm {\rm\textsf{GenGadget}} to generate $(\widetilde{k},\widetilde{\alpha},\widetilde{y},\widetilde{b},\theta^{c})$ and $|+_{\theta^{(c)}}\rangle$, and send them to the adversary. The adversary $\mathscr{A}$ has to guess the value of $\widetilde{c}$. If $c=\widetilde{c}$, then he can get the correct $|+_{\theta^{(c)}}\rangle$. However, similar to Theorem 6 in Ref. \cite{cojocaru2021possibility}, the probability of the adversary $\mathscr{A}$ guess the correct $c$ satisfies
\begin{equation}
	\label{eq:28}
	Pr[\mathscr{A}(\theta^{(c)},|+_{\theta^{(c)}}\rangle)=c]\leq \frac{1}{2} + \textsf{negl(n)},
\end{equation}
where \textsf{negl(n)} is a negligible function. Hence, the adversary $\mathscr{A}$ cannot get any useful information about $|+_\theta\rangle$. But, to generate the whole gadget, the challenger has to send the classical information about $((s_1,t_1),(s_2,t_2),...,(s_m,t_m))$ and $g(sk)$ to the adversary. As long as the CHE scheme \textsf{HE} satisfies the CPA-IND security, the adversary cannot deduce the value of $sk$ and $p[i]$ according to $g(sk)$. Therefore, the adversary also cannot get the information of $x[i],z[i]$ related to the quantum state 
\begin{equation}
	\gamma_{x, z}(g(s k))=\prod_{i=1}^{m} \textsf{X}^{x[i]} \textsf{Z}^{z[i]}(\textsf{P}^{\dagger})^{p[i]}|\Phi^{+}\rangle \langle\Phi^{+}|_{s_{i} t_{i}} \textsf{P}^{p[i]} \textsf{Z}^{z[i]} \textsf{X}^{x[i]},
\end{equation}
which means that the gadget is a maximum mixed state in the adversary's view.

\begin{theorem} \label{theorem:2}
	{\rm $\textsf{QHE}_{\textsf{CC}}$} provides q-IND-CPA secure for circuits that contain polynomially {\rm\textsf{T}} gates.
\end{theorem}
\proof For $\ell\in[0,L]$,  $\textsf{QHE}_{\textsf{CC}}^{(\ell)}$ is defined as the circuit that provides $\ell$ gadgets in the whole process. Note that if $\ell=L$, then $\textsf{QHE}_{\textsf{CC}}^{(L)} = \textsf{QHE}_{\textsf{CC}}$ and if $\ell=0$, then in $\textsf{QHE}_{\textsf{CC}}^{(0)}$, only classical evaluation keys are necessary. Based on the Lemma 1 from Ref. \cite{dulek2016quantum}, we can use the fact that for any QPT adversary interacting with $\textsf{QHE}_{\textsf{CC}}^{(\ell)}$, he only has a negligible advantage over an adversary interacting with $\textsf{QHE}_{\textsf{CC}}^{(\ell-1)}$ as 
\begin{equation}
	\begin{split}
		Pr[\textsf{PubK}^{\textsf{cpa}}_{\mathscr{A},\textsf{QHE}_{\textsf{cc}}^{(\ell)}}(\kappa)=1]&- Pr[\textsf{PubK}^{\textsf{cpa}}_{\mathscr{A},\textsf{QHE}_{\textsf{cc}}^{(\ell-1)}}(\kappa)=1] \\&\leq \textsf{negl}(\kappa).
	\end{split}
	\label{eq:31}
\end{equation}
According to Eq. (\ref{eq:31}), we can conclude that the difference between  $\textsf{QHE}_{\textsf{CC}}^{(L)}$ and $\textsf{QHE}_{\textsf{CC}}^{(0)}$ is also negligible due to
\begin{equation}
	\begin{split}
		Pr[\textsf{PubK}^{\textsf{cpa}}_{\mathscr{A},\textsf{QHE}_{\textsf{cc}}^{(L)}}(\kappa)=1]&- Pr[\textsf{PubK}^{\textsf{cpa}}_{\mathscr{A},\textsf{QHE}_{\textsf{cc}}^{(0)}}(\kappa)=1] \\&\leq \textsf{negl}(\kappa).
	\end{split}
	\label{eq:32}
\end{equation}
Since $Pr[\textsf{PubK}^{\textsf{cpa}}_{\mathscr{A},\textsf{QHE}_{\textsf{cc}}^{(0)}}(\kappa)=1] \leq \frac{1}{2}+\textsf{negl}^{'}(\kappa)$, we can get
\begin{equation}
	\begin{split}
		&Pr[\textsf{PubK}^{\textsf{cpa}}_{\mathscr{A},\textsf{QHE}_{\textsf{cc}}}(\kappa)=1]
		\\&\leq Pr[\textsf{PubK}^{\textsf{cpa}}_{\mathscr{A},\textsf{QHE}_{\textsf{cc}}^{(0)}}(\kappa)=1]+ \textsf{negl}(\kappa)
		\\&\leq\frac{1}{2}+\textsf{negl}^{'}(\kappa)+\textsf{negl}(\kappa)
	\end{split}
	\label{eq:33}
\end{equation}according to the Eq. (\ref{eq:32}).
Therefore, we can conclude that the proposed $\textsf{QHE}_{\textsf{CC}}$ satisfies q-IND-CPA. 

\begin{figure*}
	\centering
	\includegraphics[width=1\textwidth]{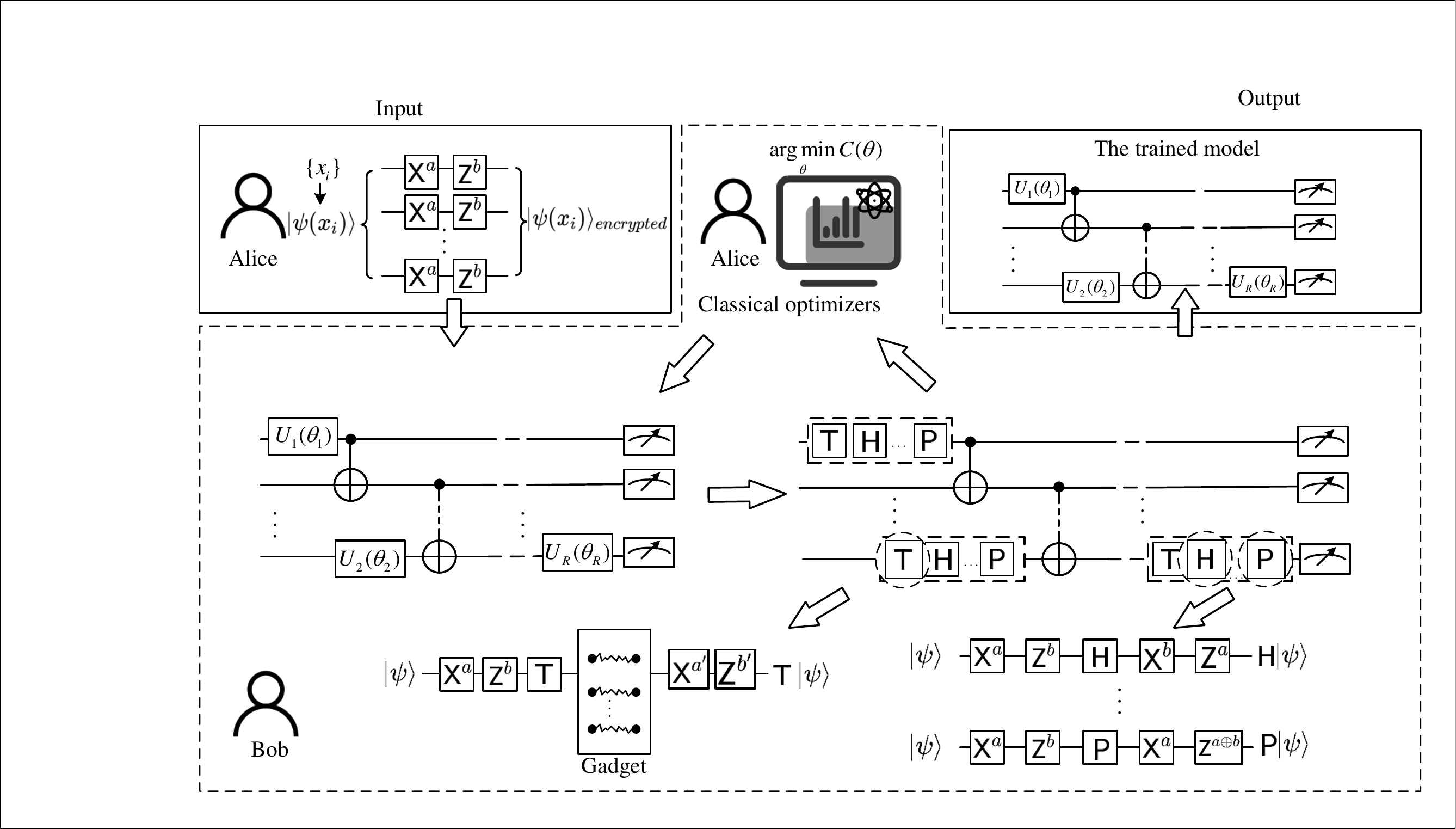}
	\caption{The process diagrams of the proposed delegated VQAs. Alice sends the encrypted quantum input states $|\psi(x_i)\rangle$ to Bob. Then Bob performs the gates in the discrete set of gates \{\textsf{X,Z,H,T,P,CNOT}\}, which are obtained by decomposing $U(\theta)$. Bob also measures the output state and sends the measurement results back to Alice. After decryption, Alice updates the parameters $\theta$ on her classical computer. Finally, Alice and Bob repeat the steps to minimize the cost function and output the trained ansatz circuit model. } \label{vqa pro2}
\end{figure*}

Next, $\textsf{QHE}_\textsf{CC}$ is compared with similar QHE schemes such as \textsf{CL} scheme \cite{broadbent2015quantum}, \textsf{AUX} scheme \cite{broadbent2015quantum}, \textsf{TP} scheme  \cite{dulek2016quantum}, and the encrypted \textsf{CNOT} scheme \cite{mahadev2020classical} in four aspects, namely the quantum capability that a client requires, the number of \textsf{T} gates that can be executed, the required universal gate set and the security of the protocol as shown in Table \ref{tab4}. The \textsf{CL} scheme \cite{broadbent2015quantum} requires the client only to perform \textsf{X,Z} gates and generate the quantum input states. But it cannot complete universal quantum computation due to \textsf{T} gates unable to be realized. In the \textsf{AUX} scheme \cite{broadbent2015quantum}, the client must perform \textsf{X, Z} gates, generate ancillary states and quantum input states. However, the client can only perform a constant number of \textsf{T} gates. In the \textsf{TP} scheme \cite{dulek2016quantum}, the client has the capability of making Bell measurements, performing $\textsf{X,Z,P}^{\dagger}$ gates and generating quantum input states. Besides, the client can perform a constant number of \textsf{T} gates. The encrypted \textsf{CNOT} scheme \cite{mahadev2020classical} allows the classical client to perform QHE, but the non-Clifford gate is \textsf{Toffoli} gate which is hard to decompose in VQA. Furthermore, in the proposed $\textsf{QHE}_\textsf{cc}$ scheme, the client only has to perform \textsf{X,Z} gates and generate quantum input states, which are the minimum requirements for a client when the input and output are quantum states. The client also can perform a constant number of \textsf{T} gates.

\section{The delegated VQAs based on $\textsf{QHE}_\textsf{CC}$} \label{sec:5}

In this section, we propose delegated VQAs based on the given QHE scheme, where a client only with the ability to  generate the quantum input qubits and perform \textsf{X,Z} gates can delegate VQAs to remote quantum servers without  disclosing his input and output.
\subsection{The proposed delegated VQAs}

Considering a situation where the client Alice owns a large database and the server Bob wishes to train a generic model with VQAs by utilizing Alice's private data such as diagnostic data of certain diseases, Alice does not want her private data to be revealed to the server in any way. It also can be described as a cooperative quantum computing between two parties, where one party provides the sensitive data and the other party provides the quantum computing power and receives the final computational model. To achieve the objective, a client-friendly delegated VQA is proposed based on $\textsf{QHE}_\textsf{CC}$, in which the client only needs to prepare input qubits, and perform Pauli \textsf{X,Z} gates. If there exists a trusty third party willing to help the client provide encrypted input qubits, the capability of the clients can even be reduced to be pure classical. Furthermore, a malicious server cannot obtain any useful information about the private data of the client.
\floatname{algorithm}{Protocol}
\begin{algorithm*}[htb]
	\caption{\textsf{The delegated VQAs based on $\textsf{QHE}_\textsf{CC}$}}
	\label{Pro 1}
	\noindent\hspace*{0.02in}{\bf Requirements:}
	The server Bob publicly announces the set of unitary operators $\{U(\theta)=\prod_{r=1}^{R} U_{r}(\theta_r)\}$ and the set of the observables $\{O_k\}$.
	
	\noindent\hspace*{0.02in}{\bf Input:}
	The client Alice provides input qubits $|\psi(x_i)\rangle$ corresponding to her classical data set $\{x_i\}$.
	
	\noindent\textbf{The preparation phase}
	\begin{algorithmic}
		\State{Step 1. The client Alice uses the Solovay-Kitaev algorithm to decomposes each $U_{r}(\theta_r)$ to a discrete gate set $U_{r}=\textsf{\{X, Z, P, H, T\}}^{\otimes n_{r}}$ on her classical computer, and records the number of \textsf{T} gates as $L$. Note that a $U_{r}(\theta_r)$ can be decomposed to $n_{r}$ gates in the set of \textsf{\{X, Z, P, H, T\}}.   }
		\State{Step 2. Based on the security parameter $\kappa$ and $L$ \textsf{T} gates, Alice uses the key generation algorithm \textsf{$\textsf{QHE}_\textsf{CC}$.\textsf{KeyGen}($1^{\kappa},1^{L}$)} to generate a series of classical secret keys $(sk_{i})^{L}_{i=0}$, public keys $(pk_{i})^{L}_{i=0}$, classical keys $(evk_{i})^{L}_{i=0}$ for quantum evaluation, and $L$ \textsf{T} gate gadgets. In addition, according to $(evk_{i})^{L}_{i=0}$, Bob could get the quantum evaluation key 
			\begin{equation}
				\rho_{evk_i}=\bigotimes^{L-1}_{i=0}(\Gamma_{pk_{i+1}}(s k_{i})\otimes|evk_{i}\rangle\langle evk_{i}|).
		\end{equation}}
	\end{algorithmic}
	
	\noindent\textbf{The computation phase}
	\begin{algorithmic}
		\State{Step 1. Suppose that Alice's data set is $\{x_{i}\}$. Then she generates $|\psi(x_i)\rangle$ and encrypts these qubits with  $\textsf{X}^{a_i} \textsf{Z}^{b_i} $, where $(a_{i},b_{i}) \in \{0,1\}$ and they are encrypted by the first public key $pk_0$. Therefore, the encrypted classical-quantum state can be described as 
			\begin{equation}
				\sum_{i=1}^{n} (\textsf{HE.Enc}_{p k_{0}}(a_i), \textsf{ HE.Enc }_{p k_{0}}(b_i)) \otimes \frac{1}{4} \textsf{X}^{a_i} \textsf{Z}^{b_i} |\psi(x_i) \rangle \langle \psi(x_i)| \textsf{Z}^{b_i} \textsf{X}^{a_i} .
		\end{equation} }Then Alice send the encrypted qubits to Bob.
		
		\State{Step 2. Bob decomposes each $U_{r}(\theta_r)$ into the product of discrete quantum gates in the set $U_{r}=\textsf{\{X, Z, P, H, T\}}^{\otimes n_{r}}$ by the Solovay-Kitaev algorithm. Then Bob applies $U_{r} \in \textsf{\{X, Z, P, H, T\}}^{\otimes n_{r}} $ on the input states. There are the following two cases to be considered.
			
			(i) If $U_{r}$ is a Clifford gate, Bob only needs to update the encrypted keys straightforwardly, since $U_{r}$ commutes with the Pauli group.
			
			(ii) If $U_{r} = \textsf{T}$, Bob should use one gadget $\Gamma_{pk_{i+1}}(sk_i)$ from the evaluation key. The specific steps are similar as homomorphic evaluatuion in $\textsf{QHE}_\textsf{CC}$.}
		
		\State{Step 3. Bob measures the observables $\{O_{k}\}$ of the output qubits and sends the measurement results to Alice. As shown in Fig. \ref{vqa measure}, Alice decrypts the results with her Pauli key and updates the parameters $\theta^{l+1}_r=\theta^{l}_r+\chi\nabla_{\theta}C(\theta)$, where $\chi$ is the learning rate, $\theta_{r}$ is the parameter of $U_{r}(\theta_r)$, $l$ means the $l$-th iteration, and the cost function $C(\theta)=\langle\psi(x_i)|U(\theta)O_k|U(\theta)^{\dagger}|\psi(x_i)\rangle.$} Note that the decryption methods are shown in Fig. \ref{vqa measure}.

	\end{algorithmic}
	\noindent\hspace*{0.02in}{\bf Output:} 
	Alice and Bob repeat all the above steps to minimize the cost function $C(\theta)$ by tuning the circuit parameters $\theta$ iteratively and finally output the trained sequence $\{\theta_1, \dots, \theta_R\}$ of the ansatz circuit $U(\theta)$.
\end{algorithm*}

Before running delegated VQAs, the server should publicly announce the set of unitary operators $\{U(\theta)=\prod_{r=1}^{R} U_{r}(\theta_r)\}$ and that of the observables $\{O_k\}$. The unitary operators $U(\theta)$, however, need to be decomposed into the product of gates in the set $\textsf{\{X,Z,P,CNOT,H,T\}}$. This task can be achieved by using the Solovay-Kitaev algorithm \cite{dawson2005solovay}, which is an efficient classical algorithm for decomposing an arbitrary single-qubit gate into a sequence of gates in a fixed and finite set. 

The proposed delegated VQAs contain the preparation phase and the computation phase. In the preparation phase, the client Alice should decompose each $U_{r}(\theta_r)$ into the product of gates in the set $\textsf{\{X, Z, P, H, T\}}$ on her classical computer and records the number of \textsf{T} gates as $L$. Then Alice prepares $L$ gadgets by using Algorithm \ref{alg:A} and generates her input qubits, encrypts these qubits and sends them to the server Bob. While in the computation phase, Bob performs the unitary operations $U(\theta)=\prod_{r=1}^{R} U_{r}(\theta_r)$ on the received encrypted states in sequence. Then he measures the observables $\{O_k\}$ of the output states and sends the measurement results to Alice. After decryption, Alice could update the parameters on her classical computer. Then Bob and Alice interact with each other to minimize the cost function $C(\theta)$ and finally get the trained circuit model as shown in Fig. \ref{vqa pro2}. The specific steps of the proposed protocol are shown as Protocol \ref{Pro 1}. 

\begin{figure}
	\centering
	\includegraphics[width=0.4\textwidth]{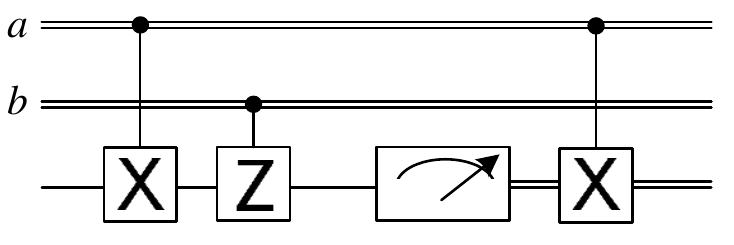}
	\caption{Decryption rules of the client Alice. For example, suppose that Alice encrypts her input qubits with the Pauli key $\textsf{X}^{a}\textsf{Z}^{b}$. If Bob measures the qubit in the computational basis and reports the result to Alice, Alice can determine the result of the corresponding measurement on her original state. The \textsf{Z} operation does not change the measurement result and the \textsf{X} operation flips it, so Alice should flip the result if $a = 1$ and do nothing if $a = 0$. } \label{vqa measure}
\end{figure}

\subsection{Correctness and security analysis}
In this part, the correctness of the computational result and the blindness of the input and output of the proposed delegated VQAs is analyzed.
\begin{theorem} \label{theorem:3}
	Correctness. If the server and client run the proposed delegated VQAs honestly, they can get the right results.
\end{theorem}
\proof In the proposed delegated VQAs, the client Alice decomposes each $U(\theta)$ into the product of gates in the set $\{\textsf{X,Z,P,H,T}\}$ by using Solovay-Kitaev algorithm \cite{dawson2005solovay}. In this algorithm, $U$ and $n$ are inputs, where $U$ is an arbitrary single-qubit quantum gate and $n$ controls
the accuracy of the approximation. The output of this algorithm is a sequence of instructions that approximates $U$ to an accuracy $\epsilon_n$, where $\epsilon_n$ is a decreasing function of $n$. As $n$ gets larger, the accuracy $\epsilon_n$ gets better. Therefore, if $n$ is good enough, the error of the ansatz circuit will have a negligible effect on the proposed delegated VQAs. Therefore if the server and client run them honestly, they can get the right results. 

\begin{theorem} \label{theorem:4}
	Blindness of the quantum input and output. For any malicious adversary, he cannot obtain any useful information about the client's input and output.
\end{theorem}
\proof The security of the client's input and output relies on $\textsf{QHE}_\textsf{CC}$. In the view of the server, each input state he received from the client is 
\begin{equation}
	\begin{split}
		\frac{1}{4} \textsf{X}^{a_i} \textsf{Z}^{b_i} |\psi_i \rangle \langle \psi_i| \textsf{Z}^{b_i} \textsf{X}^{a_i} =\frac{\mathbb{I}}{2},
	\end{split}
\end{equation}
where $|\psi_i \rangle$ is the $i$-th input qubit and $a_i,b_i$ are Pauli keys only owned by the client. Therefore, for any malicious adversary,  he also cannot obtain any useful information about the client's input as he did not know $a_i,b_i$. Similarly, the server measures the output states and sends the results to the client. However, since the server cannot get the information of Pauli keys $a_i,b_i$, he cannot know whether the classical results need to be flipped. Hence, the malicious adversary also cannot obtain any useful information about the client's output.

\section{An example of the proposed delegated VQAs and its simulation on Original Quantum Cloud} \label{sec:6}

To demonstrate the feasibility of the proposed delegated VQAs, the related usage of \textsf{T} gate gadget is simulated first  on the cloud platform of Original Quantum in order to prevent the leakage of the client's Pauli key. Then, a delegated variational quantum classifier for identifying handwritten digit images is given as an example and simulated on the cloud platform of Original Quantum. 

\subsection{An example of the T gate gadget and its simulation}
\label{sec:6.1}
\begin{figure}
	\centering
	\includegraphics[width=0.8\linewidth]{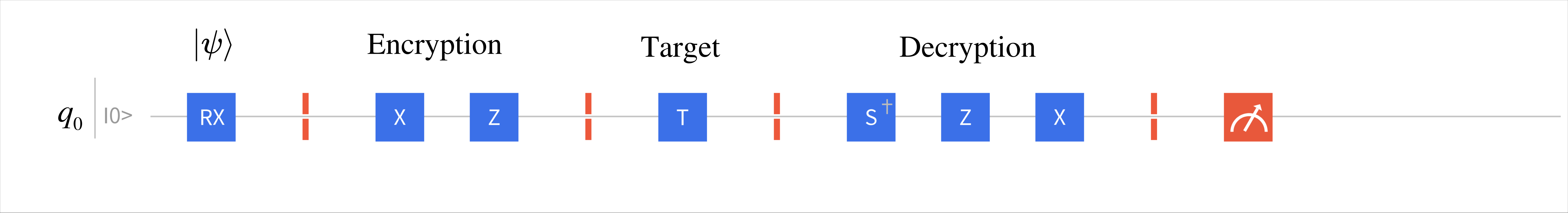}
	\caption{The quantum circuit of \textsf{T}$|\psi\rangle$, where $|\psi\rangle=R_{x}(\frac{\pi}{4})|0\rangle$ and the Pauli keys are \textsf{XZ}.} \label{circuit1}
\end{figure}

\begin{figure}
	\centering
	\includegraphics[width=0.8\linewidth]{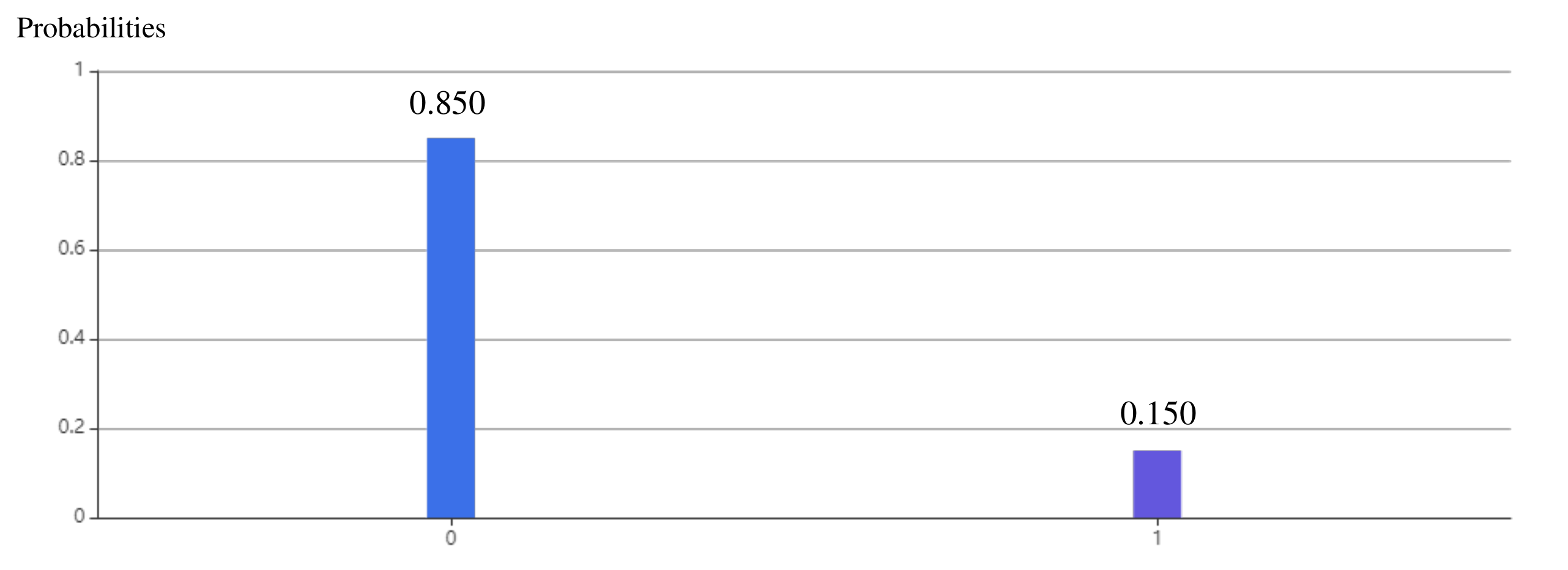}
	\caption{The results after measuring the output state of the quantum circuit of \textsf{T}$|\psi\rangle$ in Fig. \ref{circuit1} 2048 times in the $\textsf{Z}$ basis.} \label{prob1}
\end{figure}

In this subsection, we will show how to use a \textsf{T} gate gadget to correct the by-product \textsf{P} gate and simulate it on the cloud platform of Original Quantum. The key idea of the gadget is that an inverse phase gate will be applied to the qubit $\textsf{X}^a \textsf{Z}^b |\psi\rangle$ to obtain $\textsf{X}^{a^\prime} \textsf{Z}^{b^\prime} \textsf{P}^\dagger |\psi\rangle$ by using $(\textsf{P}^\dagger \otimes \textsf{I})|\Phi^{+}\rangle$ for teleportation, where $|\Phi^{+}\rangle = \frac{1}{\sqrt{2}}(|00\rangle + |11\rangle)$ and the new Pauli corrections $a^\prime, b^\prime$ depend on $a,b$ and the outcome of the Bell measurement. For example, suppose that the input state is $|\psi\rangle=R_{x}(\frac{\pi}{4})|0\rangle$, the Pauli keys are $a=1,b=1$, and the client Alice wants to perform a \textsf{T} gate on $|\psi\rangle$ to get $\textsf{T}|\psi\rangle$. Therefore, the homomorphic evaluation related to the \textsf{T} gate can be described as
\begin{equation}
	\label{eq32}
	\begin{aligned}
		\textsf{T}\textsf{XZ}|\psi\rangle=\textsf{P}\textsf{XZ}\textsf{T}|\psi\rangle,
	\end{aligned}
\end{equation}
where \textsf{XZ} are Pauli keys and there exist a \textsf{P} error needs to be corrected by performing the operation $\textsf{P}^{\dagger}$. The homomorphic evaluation circuit of Eq. (\ref{eq32}) is shown in Fig. \ref{circuit1}. After measuring the output state $2048$ times in the $\textsf{Z}$ basis, as shown in Fig. \ref{prob1}, she can get $|0\rangle$ with probability $0.850$ and $|1\rangle$ with probability $0.150$, respectively. Hence, the output state is $\textsf{T}|\psi\rangle=\sqrt{0.850}|0\rangle+\sqrt{0.150}|1\rangle$.

\begin{figure}
	\centering
	\includegraphics[width=0.35\textwidth]{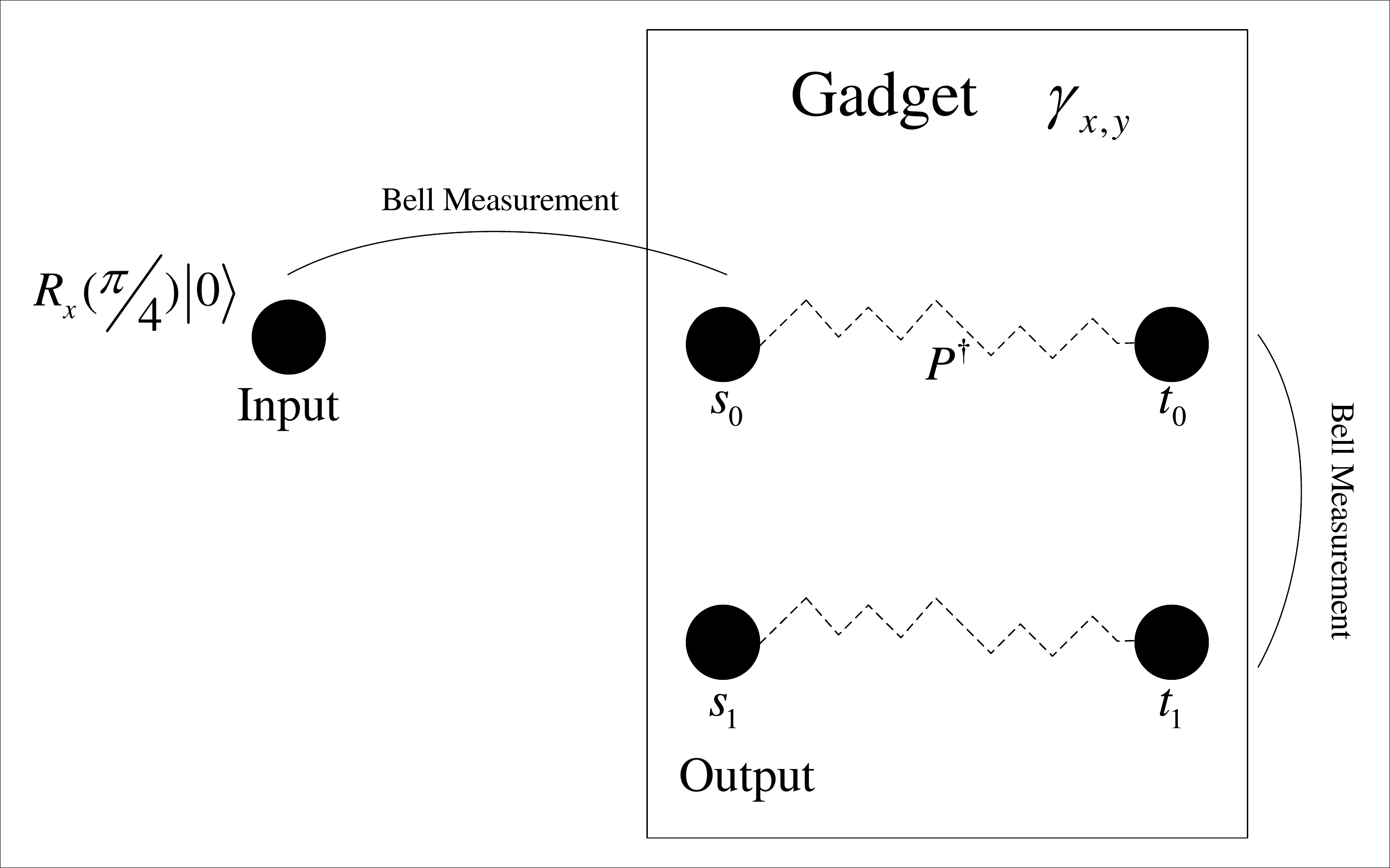}
	\caption{Schematic of the usage of the gadget. The input qubit is $|\psi\rangle=R_x(\frac{\pi}{4})|0\rangle$ and the gadget consists of two pairs of Bell states $|\Phi^{{+}}\rangle^{\otimes 2}$, the first of which is applied a $\textsf{T}^{\dagger}$ gate.} \label{sy3}
\end{figure}

\begin{figure*}
	\centering
	\includegraphics[width=0.8\linewidth]{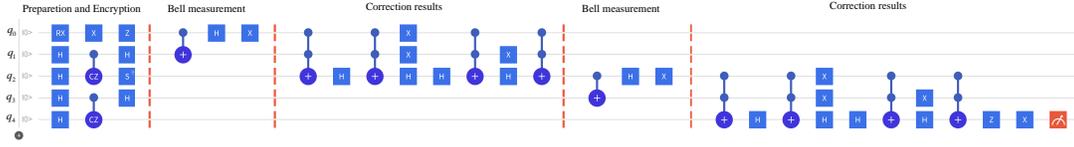}
	\caption{The quantum circuit of the usage of \textsf{T} gate gadget.} \label{circuit2}
\end{figure*}

\begin{figure}
	\centering
	\includegraphics[width=0.8\linewidth]{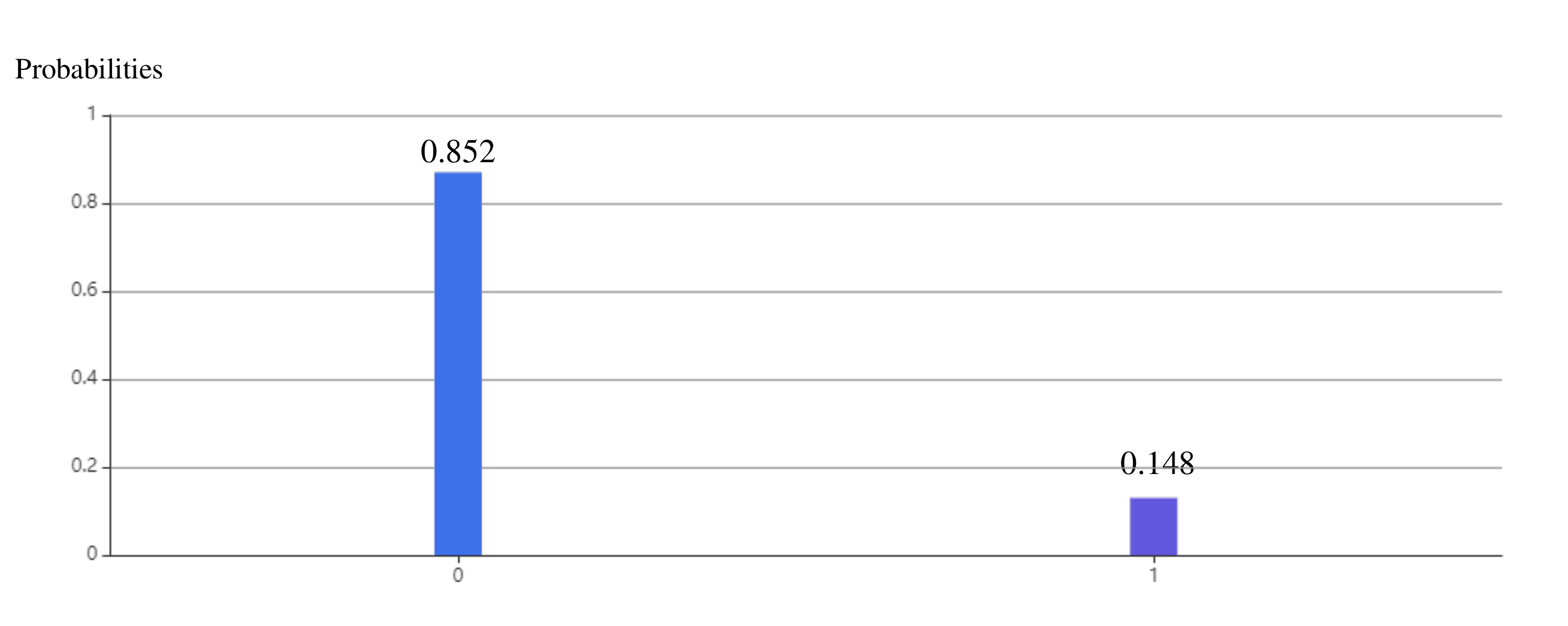}
	\caption{The measurement results in the \textsf{Z} basis of the usage of \textsf{T} gate gadget.} \label{prob2}
\end{figure}

However, in order not to reveal the Pauli key, Alice has to use the \textsf{T} gate gadget to correct the \textsf{P} error. Assume that Alice uses $\textsf{CC-RSP}_{\theta}$ \cite{cojocaru2021possibility} to generate four qubits in states $\{|+\rangle,|+_{\frac{3\pi}{2}}\rangle\}$ and corresponding classical information $((s_0,t_0),(s_1,t_1))$ to encode the Bell states which are sent to Bob. Bob performs the fixed coupling operation $\textsf{CZ}(\textsf{H} \otimes \textsf{I})$ on these qubits to get
\begin{equation}
	\begin{split}
		\textsf{CZ}(|+\rangle_{s_0} \otimes |+_{\frac{3\pi}{2}}\rangle_{t_0})\stackrel{\textsf{I} \otimes \textsf{H}}{\longrightarrow} & \frac{1}{\sqrt{2}}(|00\rangle-i|11\rangle)\\& = \textsf{X}^{0}\textsf{Z}^{0}P^{\dagger}|\Phi^{+}\rangle_{s_{0}t_{0}}, \\
		\textsf{CZ}(|+\rangle_{s_1} \otimes |+\rangle_{t_1})\stackrel{\textsf{I} \otimes \textsf{H}}{\longrightarrow} & \frac{1}{\sqrt{2}}(|00\rangle+|11\rangle)\\& = \textsf{X}^{0}\textsf{Z}^{0}|\Phi^{+}\rangle_{s_{1}t_{1}} \\
	\end{split}
\end{equation}
in terms of Eq. (\ref{eq:15}). Hence, the gadget $\gamma_{x,y}$ consists of $4$ qubits and $x[i],z[i]$ all equal $0$, where $i\in\{0,1\}$ and $p[0]=1$. According to Eq. (\ref{eq:11}), the quantum description of the \textsf{T} gadget can be defined as
\begin{equation}
	\begin{aligned}
		\gamma_{x,z} = & (X^{x[0]} Z^{z[0]}(P^{\dagger})^{p[0]}|\Phi^{{+}}\rangle\langle\Phi^{{+}}|_{s_{0}t_{0}} P^{p[0]} Z^{z[0]}X^{x[0]}) \\ &\otimes (X^{x[1]} Z^{z[1]}(P^{\dagger})^{p[1]}| \Phi^{{+}}\rangle\langle\Phi^{{+}}|_{s_{1}t_{1}} P^{p[1]} Z^{z[1]}X^{x[1]}) \\
		=&P^{\dagger}|\Phi^{{+}}\rangle\langle\Phi^{{+}}|_{s_{0}t_{0}} P \otimes |\Phi^{{+}}\rangle\langle\Phi^{{+}}|_{s_{1}t_{1}}.
	\end{aligned}
\end{equation}

Then, Bob gets a list $M$ to determine the order of measurements through an efficient classical algorithm \textsf{GenMeasurement($\widetilde{a}$)} in Ref.\cite{dulek2016quantum}. As shown in Fig. \ref{sy3}, Bob makes a Bell measurement on the input qubit in state $|\psi\rangle$ and the qubit marked as $s_{0}$, and he also makes a Bell measurement on the two qubits labeled as $t_{0}$ and $t_{1}$. The remaining qubit described as $s_{1}$ is the output qubit. Due to the uncertainty of the measurement results, some correction of the output qubit may be required. For example, the output of teleportation a $\textsf{P}^\dagger$ gate for an arbitrary single qubit $|\psi\rangle = \alpha|0\rangle+\beta|1\rangle$ can be described as
\begin{equation}
	\label{eq:9}
	\begin{aligned}
		|\psi\rangle\otimes\textsf{P}^{\dagger}|\Phi^{+}\rangle&=(\alpha|0\rangle+\beta|1\rangle)\otimes\frac{1}{\sqrt{2}}(|00\rangle+(-i)|11\rangle)\\
		&=\frac{1}{2}[|\Phi^{+}\rangle (\alpha|0\rangle+\beta(-i)|1\rangle)\\&+|\Phi^{-}\rangle(\alpha(-i)|1\rangle+\beta|0\rangle)\\
		&+|\Psi^{+}\rangle(\alpha|0\rangle-\beta(-i)|1\rangle)\\&+|\Psi^{-}\rangle(\alpha(-i)|1\rangle-\beta|0\rangle)],
	\end{aligned}
\end{equation}
where $|\Phi^{+}\rangle, |\Phi^{-}\rangle, |\Psi^{+}\rangle$ and $|\Psi^{-}\rangle$ are the Bell state. According to Eq. (\ref{eq:9}), if the measurement result is $|\Phi^{-}\rangle$, the updated Pauli corrections become $a^\prime=1, b^\prime=1$. In this case, the client applies \textsf{XZ} gate on the output qubit to obtain $\alpha(-i)|1\rangle+\beta|0\rangle \stackrel{\textsf{X,Z}}{\longrightarrow} -i(\alpha|0\rangle+\beta(-i)|1\rangle)$. The circuit of implementing the \textsf{T} gadget is shown in Fig. \ref{circuit2} and after measuring 2048 times, Alice can get the $|0\rangle$ with probability $0.852$ and $|1\rangle$ with probability $0.148$, respectively, as shown in Fig. \ref{prob2}. Thus, we can obtain the output state $|\psi\rangle_{s_0}=\sqrt{0.852}|0\rangle+|\sqrt{0.148}|1\rangle$ and it is almost identical to the ideal results without using the gadget in addition to a negligible error.

\subsection{Implementation of a delegated variational quantum classifier}

In this part, a delegated variational quantum classifier based on the variational shadow quantum learning (VSQL) for classification \cite{Li2020VSQLVS} is given. Then it is simulated on the cloud platform of Original Quantum by combing the VQNET which is a typical quantum machine learning algorithm \cite{Chen2019VQNetLF} and delegated quantum computation. The purpose of implementing delegated variational quantum classifier is to allow the cilent who only has the capabilities of performing \textsf{X,Z} gates and generating quantum input states to delegate the task of identifying handwritten digit images ``0 or 1" in the MNIST dataset \cite{LeCun2005TheMD} to the server without revealing his dataset. 
\begin{figure}[H]
	\centering
	\includegraphics[width=0.48\textwidth]{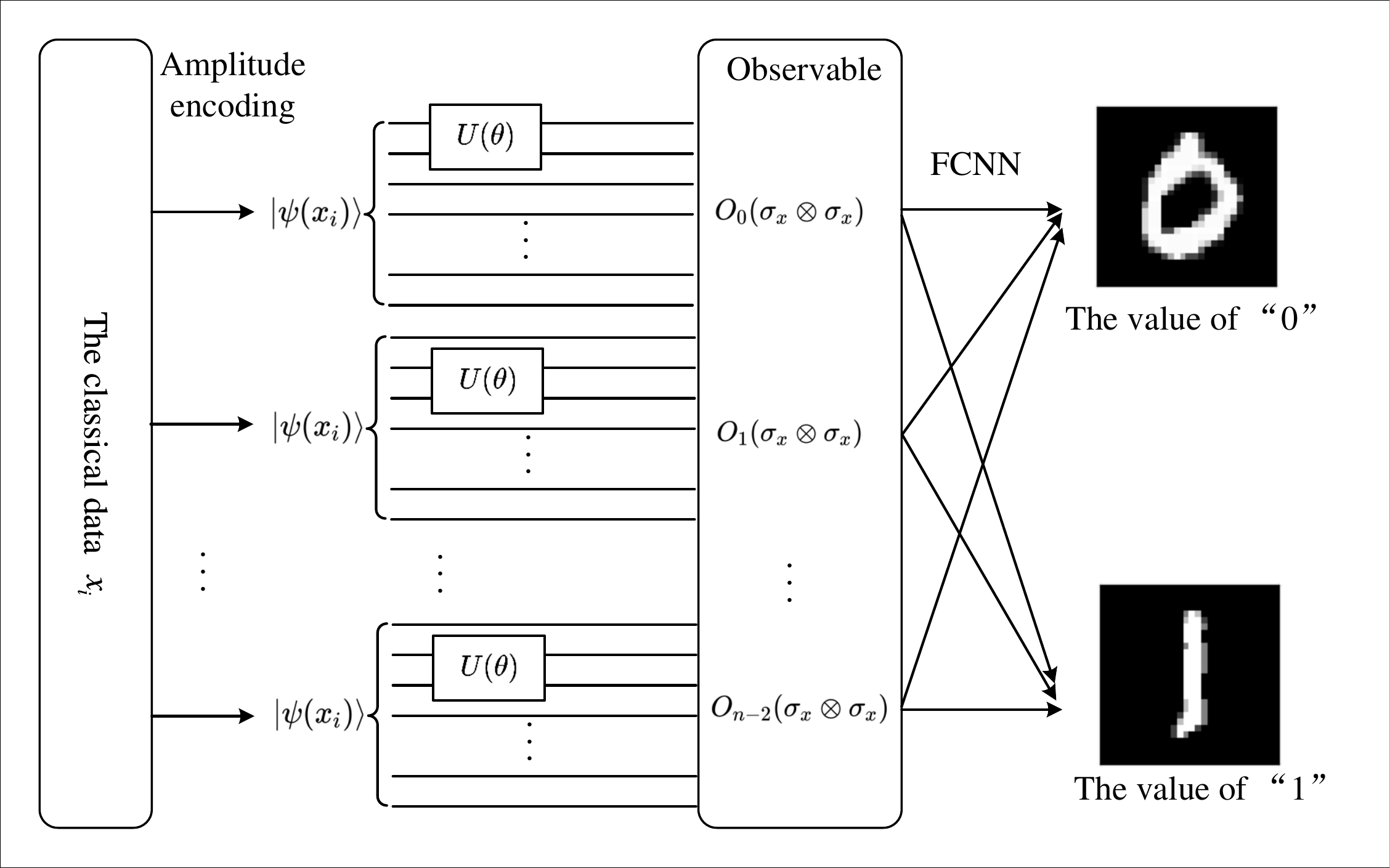}
	\caption{The model of VSQL for classification. The classical data $x_i$ is encoded as $|\psi(x_i)\rangle$ according to amplitude encoding. Then, the local parameterized quantum ansatz circuit $U(\theta)$ is applied on the input states to get the observable $O_i$. Finally, a FCNN is used to classify the handwritten digits ``0 and 1". } \label{draw}
\end{figure}
In the model of VSQL for classification \cite{Li2020VSQLVS} as shown in Fig. \ref{draw}, the classical data $x_i$ are encoded as $n$-qubit quantum states $|\psi(x_i)\rangle$ according to amplitude encoding. Then, the local parameterized quantum ansatz circuit $U(\theta)$ as shown in Fig. \ref{vqa_cir} is applied on the first two qubits and obtained the observable $O_1$ at first. Next, for the same input qubits $|\psi(x_i)\rangle$, $U(\theta)$ is applied on the second and third qubit and obtained the observable $O_2$. The similar operations are performed until $U(\theta)$ are applied to the last two qubits and the observable $O_{n-2}$ is obtained. Finally, the handwritten digits ``0 and 1" are classified based on the observables and a classical fully connected neural network (FCNN). Note that, the implementation of variational quantum classifier runs locally, so we assume that Alice ``sends" input qubits to Bob and Bob ``returns" measurement results to Alice. The details of the implementation of the delegated VSQL for classification are shown as follows.

\begin{figure}[H]
	\centering
	\includegraphics[width=0.4\textwidth]{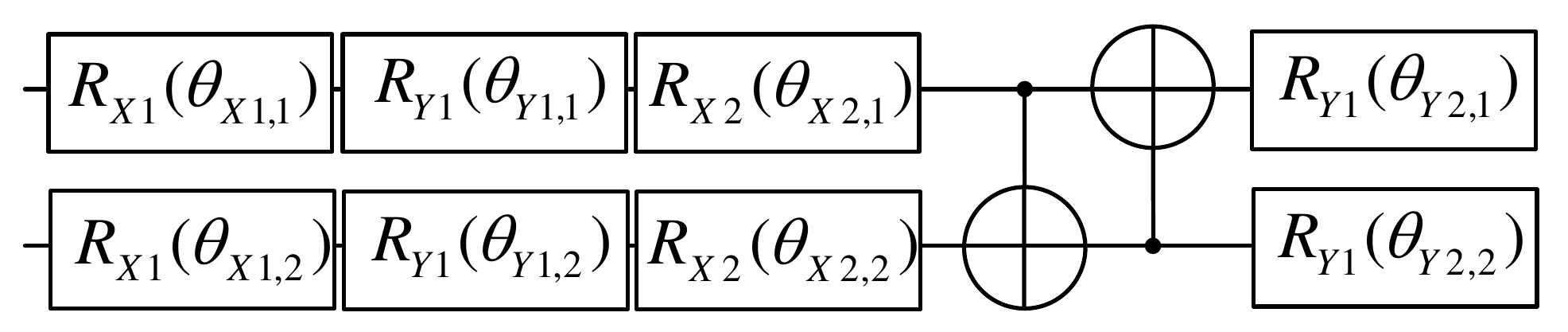}
	\caption{The local parameterized quantum ansatz circuit $U(\theta)$.} \label{vqa_cir}
\end{figure}
Before the delegated VQA starts, the server Bob publicly announces the construction of the local parameterized quantum ansatz circuit similar to that in Fig. \ref{vqa_cir}. For simplicity, we set $n=10$ as the number of input qubits, $n_{qsc}=2$ is the width of the quantum circuit and $U(\theta)$ is only applied to consecutive $n_{qsc}$ qubits each time. The $U(\theta)$ is defined as
\begin{equation}                
	\begin{split}
		U(\theta)=R_{X1}(\theta_{X1,v})R_{Y1}(\theta_{Y1,v})R_{X2}(\theta_{X2,v})\\ \textsf{CNOT}_{v-1,v}\textsf{CNOT}_{v,v-1}R_{Y2}(\theta_{Y2,v})
	\end{split}
\end{equation}
where $v$ equal to 2 is the number of input qubits and the local parameterized quantum ansatz circuit consists of two $R_x(\theta)$ parametrized by angles $ \theta_{X_1}=\{\theta_{X_1,1},\theta_{X_1,2}\}$ and $\theta_{X_2}=\{\theta_{X_2,1},\theta_{X_2,2}\}$, and two $R_y(\theta)$ parametrized by $\theta_{Y_1}=\{\theta_{Y_1,1},\theta_{Y_1,2}\}$ and $\theta_{Y_2}=\{\theta_{Y_2,1},\theta_{Y_2,2}\}$. A layer of two staggered sets of nearest-neighbor \textsf{CNOT} and the observable is $\sigma_x \otimes \sigma_x$. For the given data set $\mathcal{D}=\{\rho_{in}^{(m)},y^{(m)}\}_{m=1}^N$, the cost function is designed to be cross-entropy, which can be described as
\begin{equation}
	C(\theta, w, b ; \mathcal{D}):=-\frac{1}{N} \sum_{m=1}^N \sum_{k=1}^K y_k^{(m)} \log \hat{y}_k^{(m)}(\rho_{i n}^{(m)} ; \theta, w, b)
\end{equation}
where $w$ are weights, $b$ is the bias of the classical FCNN, and the predicted label $\widehat{y}^{(m)}$ is defined as
\begin{equation}
	\hat{y}^{(m)}\left(\rho_{i n}^{(m)} ; \theta, w, b\right)=\delta\left(\sum_i^{n-n_{qsc}+1} w_i o_i^{(m)}\left(\rho_{i n}^{(m)} ; \theta\right)+b\right),
\end{equation}
where $\delta(z)=(1+e^{-z})^{-1}$ be the sigmoid activation function and the shadow features $o_i$ is defined as 
\begin{equation}
	\begin{split}
		&o_i^{(m)}(\rho_{i n}^{(m)} ; \theta)\\=&\operatorname{Tr}(\rho_{i n}^{(m)}U^{\dagger}(\theta) (\sigma_x\otimes\sigma_x) U(\theta)).
	\end{split}
\end{equation}
Note that, in the first iteration of implementing the parameterized quantum ansatz circuit, the parameters $\theta$ of the local parameterized quantum ansatz circuit $U(\theta)$ are randomly initialized to
\begin{equation}
	\begin{split}
		&\left[\begin{array}{cccc}
			\theta_{X_1,1} & \theta_{Y_1,1} &  \theta_{X_2,1} & \theta_{Y_2,1} \\
			\theta_{X_1,2} & \theta_{Y_1,2} &  \theta_{X_2,2} & \theta_{Y_2,2}  
		\end{array}\right]  \\
		\Rightarrow
		&\left[\begin{array}{cccc}
			5.57 & 4.34 & 3.85 & 6.22\\
			5.76 & 1.40 & 5.23 & 5.05 \\ 
		\end{array}\right] .
	\end{split}
\end{equation}

\noindent \textbf{The preparation phase}

\begin{figure}
	\centering
	\includegraphics[width=0.45\textwidth]{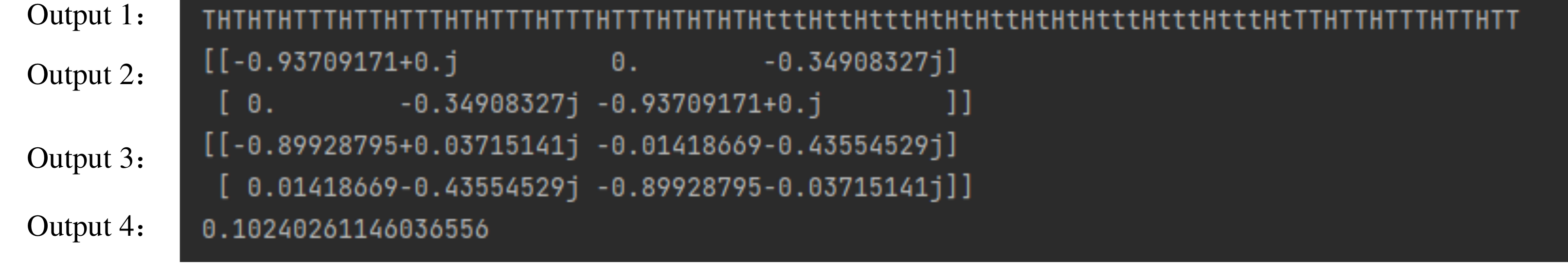}
	\caption{The decomposition of $R_x(5.57)$, where ``T" denotes a \textsf{T} gate, ``H" denotes a\textsf{H} gate and ``t" denotes a $\textsf{T}^{\dagger}$ gate. Note that Output $1$ is a sequence of decomposition of $U(\theta)$, Output $2$ and $3$ represent the matrix of $U(\theta)$ and the matrix of the decomposed sequence, respectively, and Output $4$ represents the trace distance between the two matrices.} 
	\label{decomposition}
\end{figure}

Step 1: The client Alice decomposes the gates $R_x(\theta),R_y(\theta)$ in $U(\theta)$ into a discrete gate set $\{\textsf{H,T},\textsf{T}^{\dagger}\}$ by the Solovay-Kitaev algorithm on her classical computer and records the number of \textsf{T} gates. $R_x(5.57)$ can be decomposed into $35$ \textsf{T} gates, $24$ $\textsf{T}^\dagger$ gates and $28$ \textsf{H} gates as shown in Fig. \ref{decomposition}. Therefore, for $R_x(5.57)$, Alice and Bob need to prepare $L_1=35+24$ gadgets for dealing with \textsf{T} gates and $\textsf{T}^\dagger$ gates. Note that, $\textsf{T}^\dagger$ gate can also be implemented by using the gadget. Other rotated quantum gates can be decomposed similarly and corresponding gadgets should also be prepared.

Step 2: Alice prepares $L$ gadgets in Bob's hand using the Eqs. (32-33) given in Sec. \ref{sec:6.1}.

\noindent \textbf{The computation phase}

Step 1: Suppose that Alice owns the handwritten digit images ``0 or 1" in the MNIST dataset. For each handwritten digit image which corresponds to a one-dimensional vector $x_i$, Alice generates $|\psi(x_i)\rangle$ based on the amplitude encoding. For example, if the vector $x_i=[\frac{1}{2},\frac{1}{2},\frac{-1}{2},\frac{-1}{2}]^\top$, the corresponding quantum state $|\psi(x_i)\rangle=\frac{1}{2}|00\rangle+\frac{1}{2}|01\rangle-\frac{1}{2}|10\rangle-\frac{1}{2}|11\rangle$. Alice also needs to encrypt $|\psi(x_i)\rangle$ with $\bigotimes_{i=1}^{4}\textsf{X}^{a_i}\textsf{Z}^{b_i}$ and send them to Bob. For simplicity, we set all the Pauli key $a=b=1$.

Step 2: After Bob received the input qubits, he applies the discrete quantum gates of $U(\theta)$ on these qubits similar as the homomorphic evaluation in $\textsf{QHE}_\textsf{CC}$ scheme. For example, if Bob needs to perform a \textsf{H} gate or a \textsf{CNOT} gate, he updates the Pauli keys as
\begin{equation}
	\begin{split}
		\textsf{H}&:(a=1,b=1) \rightarrow (a^{\prime}=b=1,b^{\prime}=a=1),\\
		\textsf{CNOT}_{1,2}&:(a_1=1,b_1=1,a_2=1,b_2=1) \\&\rightarrow (a_1^{\prime}=a_1=1,b_1^{\prime}=b_1 \oplus b_2=0,\\&a_2^{\prime}=a_1 \oplus a_2=0,b_2^{\prime}=b_2=1),
	\end{split}
\end{equation}
where \textsf{CNOT} is a two-qubit gate as wire $1$ is the control and wire $2$ is the target. However, if Bob needs to perform a \textsf{T} gate or a $\textsf{T}^{\dagger}$ gate, he uses a gadget and update the Pauli keys as Eq. (35) in Sec. \ref{sec:6.1}.

Step 3: Bob measures the output qubits in the \textsf{X} basis and sends the result to Alice. Alice decrypts the results with her updated Pauli keys $\textsf{X}^{a^{\prime}}\textsf{Z}^{b^{\prime}}$ as described in the decryption of $\textsf{QHE}_\textsf{CC}$ scheme and updates the parameters $\theta$ and $\{w,b\}$ based on the gradient-descent optimization method described as
\begin{equation}
	\begin{split}
		&\theta^{\prime} \leftarrow \theta-\chi\frac{\partial C}{\partial \theta},
		w^{\prime} \leftarrow w-\chi\frac{\partial C}{\partial w},
		b^{\prime} \leftarrow b-\chi\frac{\partial C}{\partial b},
	\end{split}
\end{equation}
where $\chi=0.01$ is the learning rate. Finally, Alice sends these updated parameters $\theta^{\prime},w^{\prime},b^{\prime}$ back to Bob.

\begin{figure}
	\centering
	\includegraphics[width=0.4\textwidth]{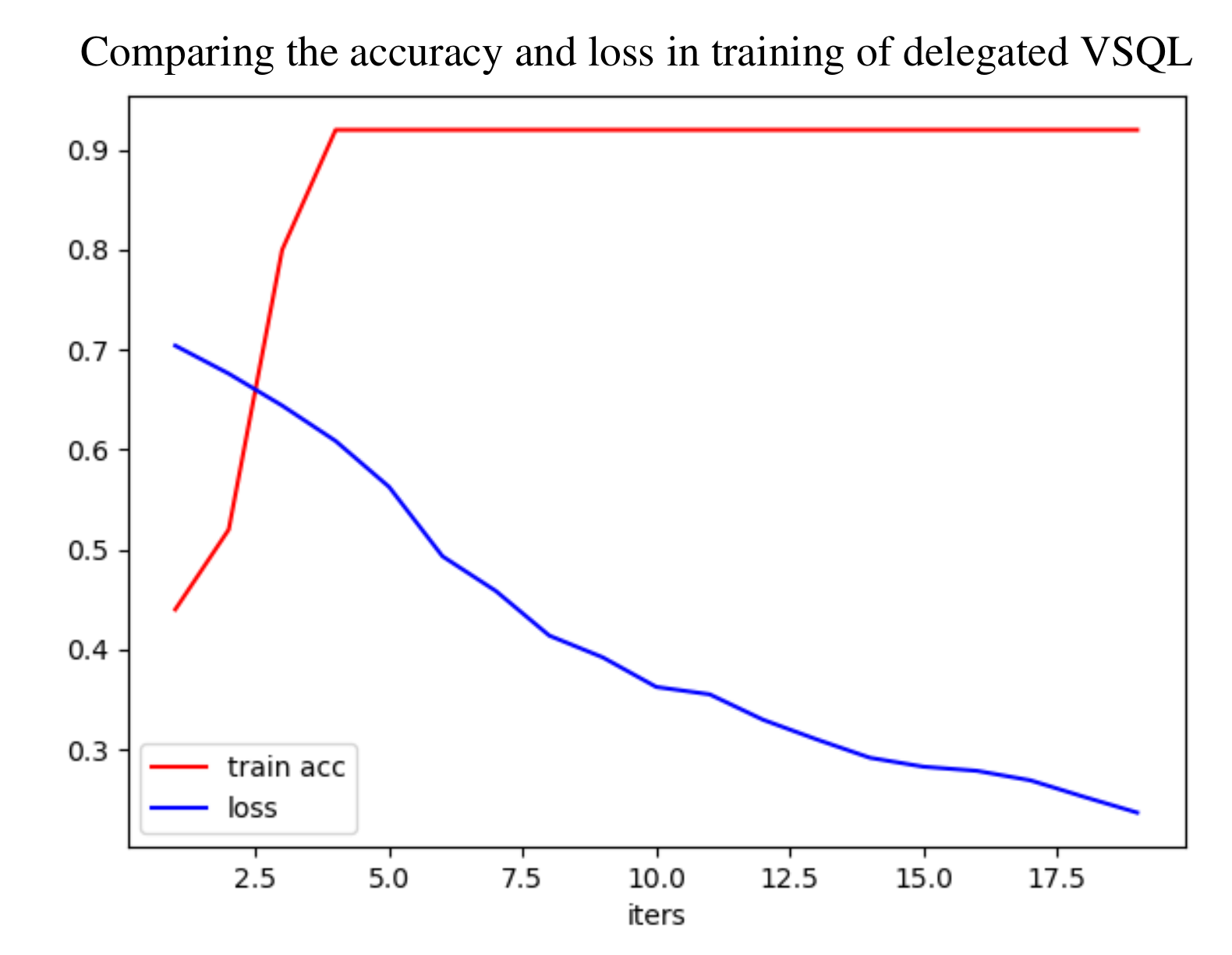}
	\caption{Comparing the accuracy and loss relationship in training of delegated VSQL for classification, where the red line is the training accuracy gradually rises to 0.94 and the blue line is the training loss, which is decreasing as accuracy increases.} \label{quantum model}
\end{figure}

They repeated the above three steps for $20$ times and the simulation result is shown in Fig. \ref{quantum model}, where the accuracy can reach 0.94. At this point, Bob can get the trained model where the parameters $\theta$ in the final iteration are 
\begin{equation}
	\begin{split}
		&\left[\begin{array}{cccc}
			\theta_{X_1,1} & \theta_{Y_1,1} &  \theta_{X_2,1} & \theta_{Y_2,1} \\
			\theta_{X_1,2} & \theta_{Y_1,2} &  \theta_{X_2,2} & \theta_{Y_2,2}  
		\end{array}\right] \\
		\Rightarrow
		&\left[\begin{array}{cccc}
			5.88 & 3.62 & 3.57 & 6.66\\
			5.78 & 0.96 & 5.87 & 5.02 \\ 
		\end{array}\right] .
	\end{split}
\end{equation}

\section{Conclusions} \label{sec:7}
We have proposed a general framework of delegated VQAs based on the improved QHE scheme $\textsf{QHE}_\textsf{CC}$, which enables the quantum server to use the client's data to train the parameterized ansatz circuit while still keeping the input data of the client private. We have analyzed the security of the proposed $\textsf{QHE}_\textsf{CC}$ and shown it can satisfy q-CPA-IND. Moreover, compared with similar QHE schemes, the requirements for the quantum capabilities of the clients are much less in the proposed $\textsf{QHE}_\textsf{CC}$ scheme since the clients only need to perform \textsf{X,Z} gates and generate quantum input states. Thus, the proposed delegated VQAs will greatly promote the application of VQAs in quantum cloud enviroments.
Besides, we have used delegated variational quantum classifier to identify handwritten digit images as a specific example of delegated VQAs and simulated it on the cloud platform of Original Quantum to show its feasibility. However, the proposed delegated VQAs is mainly suitable for various single-client and single-server tasks. How to extend it to deal with multi-party tasks such as quantum federated learning is worth further research.

\section*{Acknowledgment}
This work was supported by National Natural Science Foundation of China (Grant Nos. 62271436, U1736113, 62272483, and 61836016), the Science
 and Technology Innovation Program of Hunan Province (Grant No. 2022RC1187).	

	

\end{document}